\documentclass[a4paper,11pt]{article}
\pdfoutput=1
\usepackage{jcappub}

\usepackage{mathtools}
\usepackage[T1]{fontenc}
\newcommand{\be}{\begin{equation}}
\newcommand{\ee}{\end{equation}}
\newcommand{\rR}{\rho_r}
\newcommand{\rp}{\rho_\phi}
\newcommand{\gs}{g_\star}
\newcommand{\gss}{g_{\star s}}
\newcommand{\Tosc}{T_\text{osc}}
\newcommand{\Tqcd}{T_\text{QCD}}
\newcommand{\Tbbn}{T_\text{BBN}}
\newcommand{\Tini}{T_\text{ini}}
\newcommand{\Tend}{T_\text{fin}}
\newcommand{\Tc}{T_\text{c}}
\newcommand{\Rini}{R_\text{ini}}
\newcommand{\Rend}{R_\text{fin}}
\newcommand{\Rc}{R_\text{c}}

\newcommand{\ma}{m_a}

\title{Dark Matter Axions\\ in the Early Universe with \\a Period of Increasing Temperature}

\author[a]{Paola Arias,}
\author[b,\, c]{Nicolás Bernal,}
\author[d]{\\Jacek K. Osiński}
\author[d,\, e]{and Leszek Roszkowski}
\affiliation[a]{Departamento de Física, Universidad de Santiago de Chile\\Casilla 307, Santiago, Chile}
\affiliation[b]{New York University Abu Dhabi\\
PO Box 129188, Saadiyat Island, Abu Dhabi, United Arab Emirates}
\affiliation[c]{Centro de Investigaciones, Universidad Antonio Nariño\\Carrera 3 este \# 47A-15, Bogotá, Colombia}
\affiliation[d]{AstroCeNT, Nicolaus Copernicus Astronomical Center Polish Academy of Sciences\\ul. Rektorska 4, 00-614 Warsaw, Poland}
\affiliation[e]{National Centre for Nuclear Research\\ul. Pasteura 7, 02-093 Warsaw, Poland}
\emailAdd{paola.arias.r@usach.cl}
\emailAdd{nicolas.bernal@nyu.edu}
\emailAdd{josin@camk.edu.pl}
\emailAdd{leszek.roszkowski@ncbj.gov.pl}

\abstract{We consider the production of axion dark matter through the misalignment mechanism in the context of a nonstandard cosmological history involving early matter domination by a scalar field with a time-dependent decay rate. In cases where the temperature of the Universe experiences a temporary period of increase, Hubble friction can be restored in the evolution of the axion field, resulting in the possibility of up to three ``crossings'' of the axion mass and the Hubble expansion rate. 
This has the effect of dynamically resetting the misalignment mechanism to a new initial state for a second distinct phase of oscillation. 
The resultant axion mass required for the present dark matter relic density is never bigger than the standard-history window and can be smaller by more than three orders of magnitude, which can be probed by upcoming experiments such as ABRACADABRA, KLASH, ADMX, MADMAX, and ORGAN, targeting the axion-photon coupling. This highlights the possibility of exploring the cosmological history prior to Big Bang Nucleosynthesis through searches for axion dark matter beyond the standard window.}

\begin{document}
\begin{flushright}
    PI/UAN-2022-719FT
\end{flushright}

\maketitle

\section{Introduction}
The QCD axion emerges as an elegant solution of the so-called strong CP problem, by spontaneous breaking of the extra global $U(1)$ Peccei-Quinn symmetry at some very high scale $f_a$~\cite{Peccei:1977hh, Weinberg:1977ma, Wilczek:1977pj}. By its coupling to gluons, the corresponding pseudoscalar particle {\it i.e.} the axion, acquires a mass, which is thermally suppressed at temperatures higher than the scale of the QCD phase transition. 
Axions are natural dark matter (DM) candidates, as they have a very simple nonthermal production mechanism --  the misalignment (or realignment) mechanism -- which ensures that they are sufficiently cold and very lightweight, and thus stable on cosmological scales~\cite{Dine:1981rt}. Although it is not the only mechanism to produce cold DM axions, it is the least model dependent, provided that the Hubble parameter is much greater than the axion mass well before matter-radiation equality.  In this case, the expansion of the Universe acts as a friction term, freezing the axion amplitude to a constant value. As the temperature drops, the damping diminishes up to a point where the field can become dynamical and starts to oscillate, producing DM particles. The onset of this oscillation has been usually quoted in the literature as the ``crossing'' \cite{Marsh:2015xka}
\be \label{eq:3Hm} 
    3\, H(t) = m(t)\, ,
\ee
where $H(t)$ is the Hubble parameter and $m(t)$ is the time-dependent axion mass.\footnote{From numerical calculations, it has been found that a more accurate expression in a radiation-dominated era is $1.6\, H(t) \simeq m(t)$~\cite{Marsh:2015xka, Blinov:2019rhb}.} In the standard cosmological scenario, the (single) crossing happens during a radiation-dominated (RD) era, and the axion abundance is then set by only two parameters: its mass, and the amplitude of the field before it starts to oscillate around the minimum of its potential. 

The misalignment mechanism motivates the QCD axion as DM in the range of masses $10^{-6}$~eV $\lesssim m_a \lesssim 10^{-5}$~eV, where $m_a$ is the constant axion mass at low temperatures, leading to an intensive search for signatures in cosmology, astrophysics, stellar physics, and laboratory experiments. In addition, this has made axion physics an active research field at the theoretical and experimental forefront. The full QCD axion mass range can be explored by means of a variety of experiments that exploit different couplings of the axion to standard model (SM) particles, the most common being its coupling to two photons. Though the parameter space where the axion can explain the entire DM abundance has only barely been touched by the haloscope experiments ADMX~\cite{ADMX:2009iij, ADMX:2019uok} and HAYSTAC~\cite{HAYSTAC:2018rwy}, many prospects aim to go beyond the standard axion DM window. On the lower-mass side, which will turn out to be our focus in this work, many theoretical works have pointed out ways in which the low-mass bound can be stretched, such as entropy injection~\cite{Steinhardt:1983ia, Lazarides:1990xp, Kawasaki:1995vt, Giudice:2000ex}, nonstandard cosmological histories (NSCs)~\cite{Visinelli:2009kt, Blinov:2019rhb, Nelson:2018via, Arias:2021rer}, coupling to new hidden particles~\cite{Agrawal:2017eqm, Allali:2022yvx},\footnote{Such scenarios can be distinguished by their potential signatures to the extra fields.} a trapped misalignment mechanism~\cite{DiLuzio:2021gos, DiLuzio:2021pxd},\footnote{The parameter space where the axion can solve the DM problem ends up lying outside the QCD axion band. Therefore, such a scenario can be easily distinguished from the one analyzed here in the case of a potential future discovery.} and a stronger~\cite{Choi:1996fs}], or dynamical QCD confinement scale~\cite{Heurtier:2021rko} among others (see e.g. Ref.~\cite{DiLuzio:2020wdo}).  From the experimental side, haloscopes are not suitable for testing the region below the $\mu$eV mass range, due to their cavity size. However, several proposals aim to test this parameter space, mostly based on detecting the electric current produced by axion-induced electromagnetic fields through a pickup coil~\cite{Sikivie:2013laa, Kahn:2016aff, Ouellet:2018beu, Crisosto:2019fcj, Gramolin:2020ict}.

An interesting scenario for the dynamics of the axion in the early Universe and the misalignment production mechanism itself, is the case where the Universe goes through a period of nonadiabatic expansion during an early stage, and particularly one where the temperature temporarily increases as the Universe expands. Such a scenario has a direct impact on the axion production history which can be significantly altered compared to the misalignment mechanism in a RD Universe, depending on the details of the nonadiabatic expansion and the increasing-temperature stage.
The QCD axion is particularly sensitive to a period of increasing temperature because of its time-dependent mass. In such a case, Hubble friction can eventually be restored in the axion dynamics, damping the oscillations, and freezing the axion field amplitude for a second time.\footnote{A similar situation, in which a scalar field reenters a period of overdamping, has been studied in Ref.~\cite{Dienes:2015bka} in the context of multiple scalar fields mixing with each other and experiencing a time-dependent mass-generating phase transition.} This can happen if the crossing condition in Eq.~\eqref{eq:3Hm} occurs more than once during the thermal history. As a consequence, the initial conditions of the axion reset for a second stage of oscillations, once the temperature decreases again. 

In this work we are interested in studying axion DM production through the misalignment mechanism, assuming that the Universe experienced an Early Matter Dominated (EMD) stage before the epoch of Big Bang Nucleosynthesis (BBN), caused by a generic extra field $\phi$ that eventually decayed into SM radiation. Such periods are naturally motivated by string theory and supersymmetry, where extra-light fields are abundant and can eventually dominate the energy density in the Universe.
Without loss of generality, we parameterize the dissipation rate of \(\phi\) as $\Gamma(R,\, T) \propto R^k\, T^n$, $R$ being the scale factor and $T$ the photon temperature~\cite{Co:2020xaf, Banerjee:2022fiw}. Note that the conventional result with constant $\Gamma$ corresponds to the case where $n = k = 0$; however, in general one may expect a varying decay rate. For example, the dynamics of a coherently oscillating scalar field in the early Universe can be affected by the thermal environment due to thermal modifications of the effective potential, non-perturbative particle production, or non-topological effects~\cite{Mukaida:2012qn}. If $\phi$ oscillates with a zero-temperature mass effective potential $V(\phi) \propto m_\phi^2\, \phi^2$, it is possible to have $\Gamma\propto T$, in the limit where the effective mass (temperature corrected) is $m_\phi^\text{eff}\sim m_\phi\ll T$~\cite{Mukaida:2012qn}. In these cases, it is possible to realize $k = 0$ and $n = 1$. On the other hand, if $\phi$ oscillates in the vicinity of a potential steeper than quadratic, the decay rate also typically features a scale-factor dependence due to the field dependence of its mass~\cite{Garcia:2020wiy, Bernal:2022wck}. This can lead to a variety of values of $k$  depending on the spin of the decay products\footnote{Note that $\Gamma \propto m_\phi$ or $1/m_\phi$ for fermions or bosons in the final state, respectively.} and the shape of the potential of $\phi$ during reheating. For example, for reheating in a quartic potential, one has $\Gamma \propto R^{3/4}$ $(R^{-3/4})$ if $\phi$ decays to bosons (fermions)~\cite{Garcia:2020wiy, Bernal:2022wck}; this corresponds to $n = 0$ and $k = 3/4$, or $n = 0$ and $k = -3/4$. We consider a general parametrization for the decay rate which captures a variety of dynamics during reheating, including the effect of $i)$ decays via higher-order operators~\cite{Co:2020xaf, Moroi:2020has}, $ii)$ shapes of potentials of $\phi$ during reheating~\cite{Daido:2017wwb, Garcia:2020wiy, Bernal:2022wck}, as well as $iii)$ feedback from the thermal background~\cite{Mukaida:2012qn, Mukaida:2015nos}. We reiterate that in this work we will remain quite general when it comes to the field \(\phi\) in order to study the full range of effects caused by the modified cosmology itself. Upon specializing to a specific model, one may find that the final results are not as pronounced as the maximum extent we present. However, we find it meaningful to analyze the effects of a modified history as a guide to future work.

Our manuscript is organized as follows: in Section~\ref{sec:standard} we briefly review the misalignment mechanism in a RD Universe and define what is the hinted parameter space for axion DM. In Section~\ref{sec:NSC} we give the details of the non-standard cosmology to be considered and its main features. In Sections~\ref{sec:analytic} and~\ref{sec:results} we analyze the axion DM production through the misalignment mechanism during the period of NSC first analytically and then numerically, respectively.
In Section~\ref{sec:photons} we map our results onto the exclusion plot for the axion coupling to two photons, 
and mention some of the new generation of experiments that will be able to probe the extended parameter space. Finally, in Section~\ref{sec:concl} we summarize and conclude.

\section{Misalignment with Standard Cosmology}
\label{sec:standard}
In the standard cosmological scenario, the Universe was radiation dominated throughout the period after inflationary reheating and before BBN. The production of DM axions must be set before matter-radiation equality and one of the most natural ways to accomplish this is through the misalignment mechanism~\cite{Dine:1982ah, Arias:2012az}.

This mechanism is quite generic for low-mass scalar fields and relies on the assumption that fields in the early Universe have a random initial state, fixed by the expansion of the Universe. For concreteness, let us write down the axion Lagrangian density as 
\be \label{axion_lag}
    \mathcal L = \frac12\, \partial_\mu a\, \partial^\mu a - m^2(t)\, f_a^2 \left(1-\cos\frac{a}{f_a}\right),
\ee
where $a$ is the axion field. The potential energy above is a parameterization of the non-perturbative QCD effects that stabilize the axion at $\langle a/f_a\rangle=0$. Therefore, the axion mass turns out to be dependent on the topological susceptibility of QCD, $\chi(T)$, as
\be 
    m^2(T) = \frac{\chi(T)}{f_a^2}\,,
    \label{eq:thermal_mass1}
\ee
where $\chi(T)$ has been estimated from lattice QCD simulations and found to have a zero-temperature value of $\chi_0  \equiv \chi(0) \simeq 0.0245$~fm$^{-4}$, in the symmetric isospin case~\cite{Borsanyi:2016ksw}. However, an analytical estimate that fits the simulation quite well is given by
\begin{equation}
    m(T) \simeq m_a \times
    \begin{dcases}
    (\Tqcd/T)^4 & \text{ for } T \geq \Tqcd\,,\\
    1 & \text{ for } \Tqcd \geq T\,,
    \end{dcases}
    \label{eq:thermal_mass2}
\end{equation}
with $\Tqcd \simeq 150$~MeV and $m_a$ the zero-temperature axion mass, which from Eq.~\eqref{eq:thermal_mass1} can be expressed in terms of the decay constant $f_a$ as
\be\label{eq:ma_fa}
    m_a\approx 5.69~\mbox{meV} \left( \frac{10^9\, \mbox{GeV}}{f_a}\right).
\ee
For our estimates, we will use the approximation in Eq.~\eqref{eq:thermal_mass2}. 

Now, we move to find the evolution of the  zero mode which is given by
\be \label{eq:axion_eom}
    \ddot \theta + 3\, H(t)\, \dot\theta + m^2(t)\, \sin \theta = 0\,,
\ee
where $\theta\equiv a(t)/f_a$, and the Hubble expansion rate in a RD Universe is
\begin{equation}
    H_r(T) = \frac{\pi}{3} \sqrt{\frac{\gs(T)}{10}} \frac{T^2}{M_{\rm P}},
\end{equation}
with $\gs(T)$ the relativistic degrees of freedom contributing to radiation and $M_{\rm P}$ the reduced Planck mass. From Eq.~\eqref{eq:axion_eom} one can distinguish two regimes: in the first one, when  $3\, H(t)\gg m(t)$,  the oscillation is overdamped and the amplitude is frozen. Later, the Hubble friction decreases and the axion starts to oscillate at $t = t_{\rm osc}$, given by Eq.~\eqref{eq:3Hm}. From here,  the field can start rolling down the potential and oscillate around $\langle \theta\rangle=0$. Within the WKB approximation and assuming an adiabatic evolution,\footnote{Therefore, assuming on the one hand $\theta\ll 1$ and on the other hand, a slow variation on time of $H$ and $m(T)$.}
it is found that the amplitude of the field evolves as
\be
    a(t) \simeq a_{\rm i} \left[\frac{m_{\rm osc}}{m(t)}\, \left(\frac{R_{\rm osc}}{R}\right)^3\right]^{1/2}\, \cos \left[\int m(t')\, dt'\right],
\ee
where we assume $a_{\rm i}\sim a(t_{\rm osc})$, $R_{\rm osc}\equiv R(t_{\rm osc})$, and $m_\text{osc} \equiv m(t_\text{osc})$. The solution corresponds to fast oscillations with a slow amplitude decay; therefore, the energy density stored in the field can be expressed as \be
\label{eq:rho_a_t}
    \rho_a(t) = \frac{\dot a^2}{2} + m^2(t)\, f_a^2 \left(1-\cos\frac{a}{f_a}\right)
    \simeq \frac{ a_{\rm i}^2\, m(t)\, m_{\rm osc}}{2} \left(\frac{R_{\rm osc}}{R(t)}\right)^3\,.
\ee
The energy density at present, for nonrelativistic axions can be found in terms of temperatures by using the conservation of comoving entropy $S= s\,R^3=2\pi \gss(T) T^3\,R^3/45$,  where the function $\gss$ tracks the effective number of relativistic degrees of freedom present in the SM entropy \cite{Drees:2015exa}, as
\be
    \rho_a(T_0) = \frac{\theta_{\rm i}^2\, f_a^2} 2\, m_a\, m_{\rm osc} \frac{\gss(T_0)}{\gss(\Tosc)} \left(\frac{T_0}{\Tosc}\right)^3.
\ee
The temperature $\Tosc$ at the time of onset of oscillations can be found as
\be \label{eq:Tosc_std}
    \Tosc \simeq
    \begin{dcases}
        \left( \gs(\Tosc)^{-1/2}\, \ma\, M_{\rm P}\right)^{1/2} &\text{for }\,\Tosc\leq \Tqcd\,,\\
        \left(\gs(\Tosc)^{-1/2}\, \ma\, M_{\rm P}\, \Tqcd^4\right)^{1/6} &\text{for }\,\Tosc\geq \Tqcd\,.
    \end{dcases}
\ee
Therefore, the relic density for nonrelativistic axions produced by the misalignment mechanism is found to be
\be
\Omega h^2\approx
    \begin{dcases}
        0.003 \left(\frac{\theta_{\rm i}}{1}\right)^{2}\left(\frac{\ma}{5.6~\mu \mbox{eV}}\right)^{-3/2} &\text{ for } \ma \lesssim 3\, H(T_{\rm QCD})\,,\\
        0.08 \left(\frac{\theta_{\rm i}}{1}\right)^{2}\left(\frac{\ma}{5.6~\mu \mbox{eV}}\right)^{-7/6} &\text{ for } \ma \gtrsim 3\, H(T_{\rm QCD}).
    \end{dcases}
    \label{eq:relic_stdd}
\ee
The initial value of the misalignment angle $\theta_{\rm i}$ is fixed to $\pi/\sqrt{3}$ if the PQ symmetry breaks after inflation or is randomly chosen between $\left[-\pi,\, \pi\right]$ in the scenario where it breaks before inflation. 
In the first case, there is also good theoretical motivation to assume that DM axions are also sourced by topological defects, in a quantity still in dispute, presumably of the same order as the misalignment one~\cite{Harari:1987us, Hagmann:2000ja, Wantz:2009it, Hiramatsu:2010yu, Kawasaki:2014sqa, Gorghetto:2018ocs}.
For our purposes, we consider only the misalignment production and in order to account for both pre- and post-inflation scenarios we will define the so-called ``axion DM window'' by considering $\theta_{\rm i}$ in the range $\left[1/2,\, \pi/\sqrt{3}\right]$, which leads to axion masses in the approximate range of $56~\mu\mbox{eV}\lesssim \ma\lesssim 5$~meV. It is worth commenting at this point that for angles $\theta_{\rm i} \gtrsim 1$ the analytical result for the relic density above deviates significantly from the numerical result, as it does not account for anharmonic terms in the axion potential \cite{Lyth:1991ub, Turner:1985si, Bae:2008ue,Karamitros:2021nxi}. Nevertheless, we will not include the anharmonic corrections to the relic density, as they are not needed for our arguments, but will just comment when necessary.  

\section{Nonstandard Cosmology with a time-dependent Decay Width} \label{sec:NSC}
In general, the energy density of the post-inflationary Universe prior to BBN can be dominated by something other than radiation, resulting in a period of expansion that deviates from standard cosmology~\cite{Allahverdi:2020bys}. Such a period can have important consequences for early-Universe processes, including DM production, and will modify the discussion presented above~\cite{Steinhardt:1983ia, Lazarides:1990xp, Kawasaki:1995vt, Giudice:2000ex, Grin:2007yg, Visinelli:2009kt, Nelson:2018via, Visinelli:2018wza, Ramberg:2019dgi, Blinov:2019jqc, Carenza:2021ebx, Venegas:2021wwm, Bernal:2021yyb, Arias:2021rer, Bernal:2021bbv}. It is generally assumed that the temperature of the Universe always decreases in such phases; however, as shown in Refs.~\cite{Co:2020xaf, Ahmed:2021fvt, Barman:2022tzk, Banerjee:2022fiw} there are well-motivated scenarios in which the temperature can remain constant or even increase for an extended period of time. This allows the thermal background to pass through the same temperature multiple times, with different values of the Hubble expansion rate.

The background evolution is given by the following Boltzmann equations
\begin{align} \label{eq:boltzmann1}
    \frac{d\rp}{dt} + 3\, H\, \rp = - \Gamma(T,\, R)\, \rp\,, \\
    \frac{d\rR}{dt} + 4\, H\, \rR = + \Gamma(T,\, R)\, \rp\,, \label{eq:boltzmann2}
\end{align}
where $\rR$ and $\rp$ denote the SM radiation and the NSC-driving field energy densities, respectively. The Hubble expansion rate $H$ is given by
\begin{equation}
    H \equiv \sqrt{\frac{\rp + \rR}{3\, M_{\rm P}^2}}\,,
\end{equation}
with
\begin{equation}
    \rR(T) = \frac{\pi^2}{30}\, \gs(T)\, T^4.
\end{equation}

We assume a NSC of matter domination (by the $\phi$ field) during the period defined by the range $\Rini\leq R\leq \Rend$, which returns to a RD Universe for $R\geq \Rend$. The scale factors at the beginning and end of the NSC are associated with the initial and final temperatures of the period, $\Tini$ and $\Tend$, respectively. 

The time-dependent decay rate appearing in the Boltzmann equations can be parameterized as
\begin{equation}
    \Gamma(T,\, R) = C \left(\frac{T}{\Tend}\right)^n \left(\frac{R}{\Rend}\right)^k H(\Rend)\,,
\end{equation}
where $C$ is a parameter of order one~\cite{Barman:2022tzk}. Standard perturbative decay of $\phi$ with a constant decay width is recovered for $n = k = 0$.
Here, we focus on the case where $n < 4$ and $n - k < 5/2$, which guarantees efficient energy transfer from $\phi$ to SM radiation~\cite{Co:2020xaf}.
It is worth mentioning that if $k = 3/2$, the SM temperature is constant in the nonadiabatic phase.
Additionally, $k < 3/2$ ($k>3/2$) induces a decrease (increase) of the temperature during the nonadiabatic phase.

Recalling the features of a matter-dominated NSC, it is characterized by an adiabatic phase (for $\Rini < R < \Rc$) where the temperature cools according to $T \propto R^{-1}$ as in standard RD, followed by a nonadiabatic phase (for $\Rc < R < \Rend$) in which decays of $\phi$ modify the temperature relation. The transition between the two regimes occurs when \(R \approx \Rc\). 
As a function of the scale factor, the Hubble expansion rate can be estimated to be
\begin{equation} \label{eq:Hubble}
    H(R) \simeq
    \begin{dcases}
        H_r(\Tend) \left(\frac{\Rend}{\Rini}\right)^{3/2} \left(\frac{\Rini}{R}\right)^2 &\text{ for } R \leq \Rini\,,\\
        H_r(\Tend) \left(\frac{\Rend}{R}\right)^{3/2} &\text{ for } \Rini \leq R \leq \Rend\,,\\
        H_r(\Tend) \left(\frac{\Rend}{R}\right)^2 &\text{ for } \Rend \leq R\,.
    \end{dcases}
\end{equation}
By analytically solving the system of Eqs.~\eqref{eq:boltzmann1} and~\eqref{eq:boltzmann2}~\cite{Co:2020xaf, Barman:2022tzk}, it can be shown that the SM temperature scaling is given by
\begin{equation} \label{T(a)}
    T(R) \simeq
    \begin{dcases}
        \Tc\, \frac{\Rc}{R} &\text{ for } R \leq \Rc\,,\\
        \Tc \left(\frac{\Rc}{R}\right)^\frac{3+2x}{8} &\text{ for } \Rc \leq R \leq \Rend\,,\\
        \Tend\, \frac{\Rend}{R} &\text{ for } \Rend < R\,,
    \end{dcases}
\end{equation}
with $\Tc$ the temperature at which the decay of $\phi$ starts to affect the temperature evolution, and 
\begin{equation}
    x \equiv \frac{3\, n - 8\, k}{2\, (4-n)}\,,
\end{equation}
such that the case of constant decay rate is reproduced when $ x = 0$. An increase in temperature is obtained during the nonadiabatic phase for $x<-3/2$, in which case \(\Tc\), rather than \(\Tend\), corresponds to the minimum temperature reached during the NSC.\footnote{As discussed in Ref.~\cite{Co:2020xaf}, the dynamics of a scalar field such as rotations and fluctuations in a quadratic potential can lead to a dissipation rate \(\Gamma \propto T^3/\phi^2\), where \(x = -15/2\), or \(\Gamma \propto m_\phi^3/\phi^2\), where \(x = -3\), depending on whether the temperature or mass is relevant. The case where \(x = -3/2\) instead corresponds to a period of constant temperature, which can be achieved by a coherently oscillating scalar field with dissipation \(\Gamma \propto T^2/\phi\) or \(\Gamma \propto m_\phi^2/\phi\).}
Thus, the NSC is fully characterized by {\it three} independent parameters. A convenient choice -- which we will adopt in what follows --  is $x$, $\Tc$, and $\Tend$. The initial NSC temperature \(\Tini\) is then uniquely determined once values for these three parameters are chosen. Moreover, if one specializes to a particular model that realizes the time-dependent dissipation of \(\phi\), our three chosen parameters can be equivalently exchanged for the relevant parameters of the model.
We emphasize here that once the latter three parameters are fixed, the whole cosmological history of the field $\phi$ is completely determined. For example, before decaying, its energy density is given by $\rp(R) \simeq \rR(\Tend)\, (\Rend/R)^3$.
In addition, the ratio of energy densities $\rp/\rR$ well before the decay is given by
\begin{equation}
    \frac{\rp}{\rR}(T) \simeq \frac{\Tc}{T} \left(\frac{\Tc}{\Tend}\right)^\frac{12 - 8x}{3 + 2x},
\end{equation}
for $T > \Tini$, with which one can determine the ratio at the end of inflationary reheating, when $T = T_\text{rh}$.
As one could expect $T_\text{rh} \sim \mathcal{O}\left(10^{13}\right)$~GeV, even a completely subdominant population of $\phi$ could generate the required nonstandard cosmological era.

In terms of $x$, the decay rate of $\phi$ can be cast as
\begin{equation}
    \Gamma = \Gamma(R) = C \left(\frac{\Rend}{R}\right)^x H_{\rm fin}
\end{equation}
during the nonadiabatic phase $\Rc < R < \Rend$.
Finally, to avoid trouble with BBN, we require both $\Tend$ and $\Tc$ to be higher than $\Tbbn \simeq 4$~MeV~\cite{Kawasaki:1999na, Kawasaki:2000en, deSalas:2015glj, Hasegawa:2019jsa}.

\begin{figure}
    \def\sepf{0.99}
	\centering
    \includegraphics[scale=\sepf, trim = 0.2cm 0cm 0.8cm 0cm, clip = true]{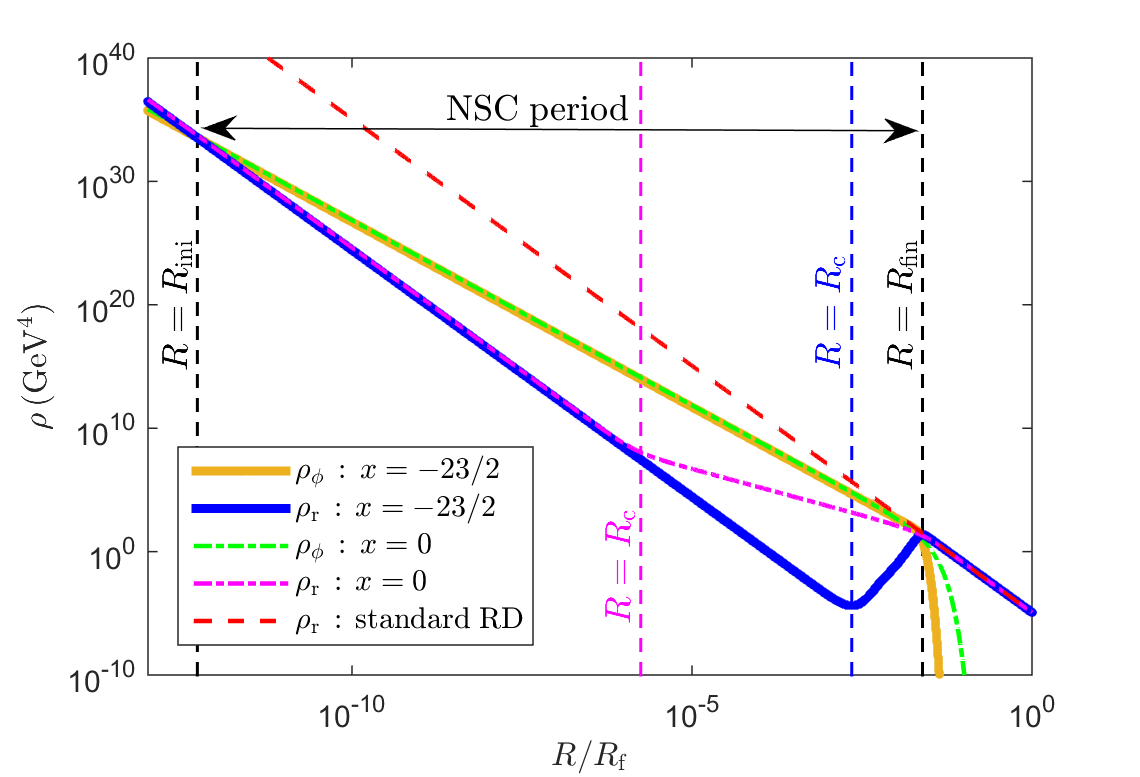}
    \caption{Evolution of the SM radiation and $\phi$ energy densities, $\rR$ and $\rp$,  
    as a function of the scale factor $R$ in three different histories.
    The red dashed line corresponds to the standard cosmological evolution of $\rR$, whereas the thin dash-dotted and the thick solid lines show two examples of NSC: $x = 0$ with $\Tc = 50$~GeV, and $x = -23/2$ with $\Tc = 54$~MeV, respectively, where $\Tend = 1$~GeV for both.
    The vertical lines show $R = \Rini$, $R = \Rc$, and $R = \Rend$. All lines are normalized to an arbitrary value of the scale factor, \(R = R_{\rm f}\), occurring in RD well after the decay of \(\phi\). 
    }
	\label{fig:bkg_rho}
\end{figure}

\begin{figure}
    \def\sepf{0.60}
	\centering
    \includegraphics[scale=\sepf, trim = 0.2cm 0cm 0.8cm 0cm, clip = true]{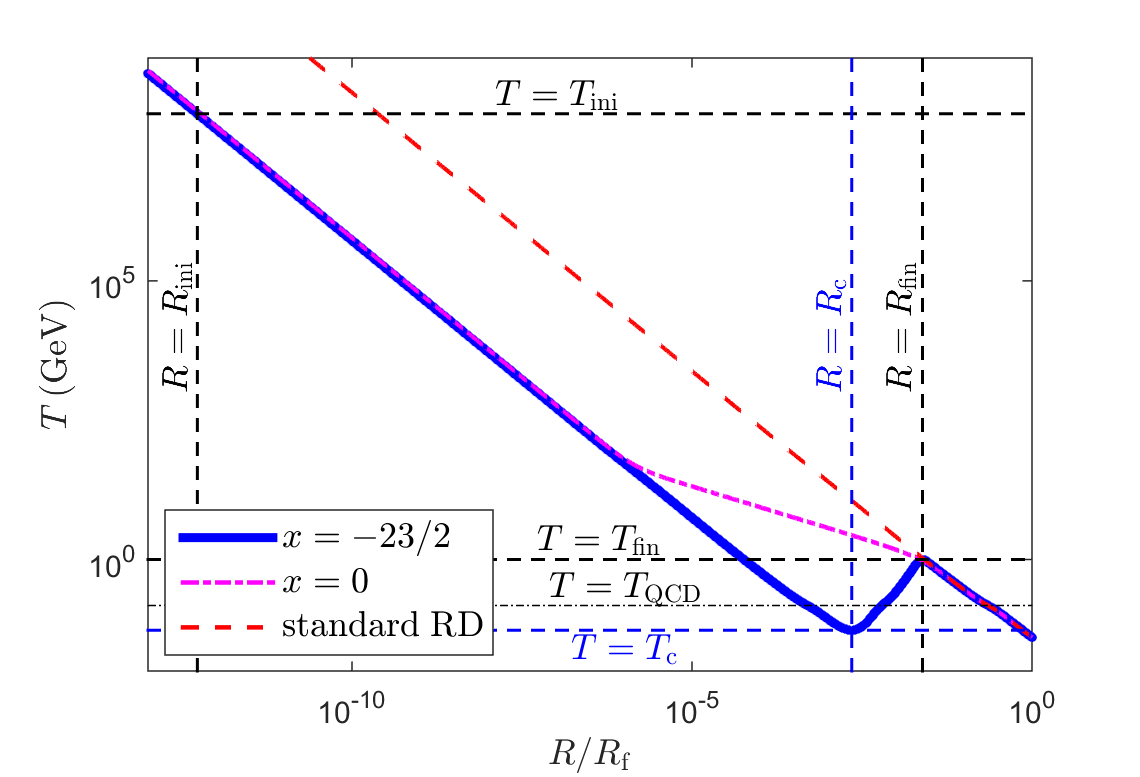}
    \includegraphics[scale=\sepf, trim = 0.2cm 0cm 0.8cm 0cm, clip = true]{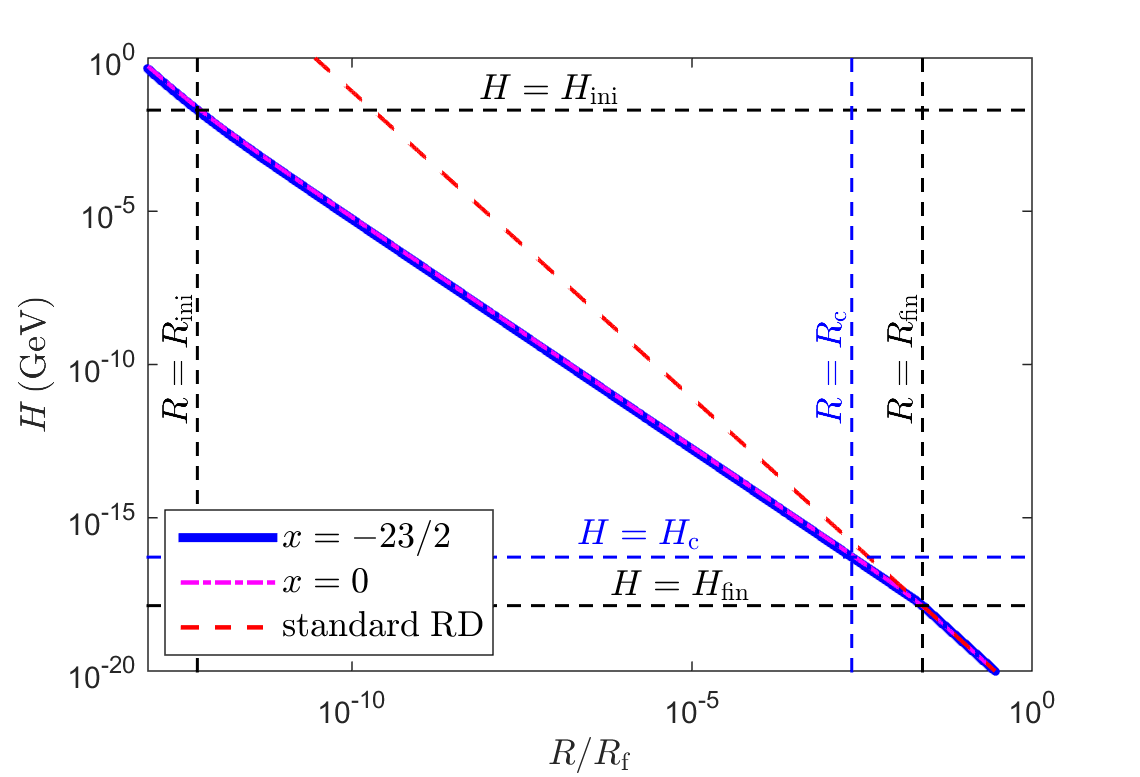}
    \caption{Evolution of the SM temperature (left) and Hubble expansion rate (right) for the histories shown in Fig.~\ref{fig:bkg_rho}.
    The red dashed lines correspond to the standard cosmological scenario, whereas the thin dash-dotted and the thick solid lines to examples of NSC with $x = 0$ and $\Tc = 50$~GeV, or $x = -23/2$ and $\Tc = 54$~MeV, respectively, with $\Tend = 1$~GeV. The lines for $\Rc$, $\Tc$, and $H_{\rm c}$ are shown only for the $x = -23/2$ case. 
    }
	\label{fig:bkg2}
\end{figure}

In Fig.~\ref{fig:bkg_rho} we show the evolution of the SM radiation and $\phi$ energy densities as a function of the scale factor $R$ for three different histories.
The red dashed line corresponds to the standard cosmological evolution of $\rR$ (with no line for $\rp$), whereas the thin dash-dotted and the thick solid lines correspond to two examples of NSC with $x = 0$ and $\Tc = 50$~GeV ({\it i.e.} standard EMD era), or $x = -23/2$ and $\Tc = 54$~MeV (coming {\it i.e.} from $n = 3$ and $k = 4$), respectively, with $\Tend = 1$~GeV for both examples.
These curves come from the full numerical solution of the Boltzmann equations in Eq.~\eqref{eq:boltzmann1} and~\eqref{eq:boltzmann2}, with the temperature dependence of the degrees of freedom \(\gs(T)\) taken into account using data from Ref.~\cite{Borsanyi:2016ksw}.
Additionally, the vertical lines show $R = \Rini$, $R = \Rc$, and $R = \Rend$.
We emphasize that the NSC era occurs between $\Rini < R < \Rend$, where $\Rini < R < \Rc$ and $\Rc < R < \Rend$ correspond to the adiabatic and nonadiabatic phases, respectively. Note that we normalize the scale factor to an arbitrary value \(R = R_{\rm f}\) occurring well after the end of NSC, which corresponds to a temperature \(T_{\rm f} < \Tc\) such that all three thermal histories coincide once standard RD is established.
The corresponding evolution of the SM temperature and the Hubble expansion rate is shown in Fig.~\ref{fig:bkg2}.
It can be seen that in the nonadiabatic phase the SM temperature decreases more slowly than in the standard case, and can even increase as in the case where $x = -23/2$.
Concerning $H$, however, the evolution in the two cases is identical, as it does not depend on the temperature but only on the total energy density of the Universe. As any one of the three parameters $x$, $\Tc$, or $\Tend$ is varied with the others held fixed, the start time of the NSC period will change, as will the initial value of the \(\phi\) energy density at any given time before NSC. Thus variation of the initial conditions is automatically accounted for by the variation of our three chosen parameters. 

The decay of $\phi$ injects entropy to the SM bath, and dilutes any preexisting DM population.
The dilution factor is defined as the ratio of the SM entropy after and before the decay, and can be estimated as follows
\begin{equation}
    \frac{S(R)}{S(\Rend)} \simeq
    \begin{dcases}
        \left(\frac{\Tend}{\Tc}\right)^\frac{15 - 6x}{3 + 2x} &\text{ for } R \leq \Rc\,,\\
        \left(\frac{\Tend}{T}\right)^\frac{15-6x}{3+2x} &\text{ for } \Rc \leq R \leq \Rend\,,\\
        1 &\text{ for } \Rend \leq R\,,
    \end{dcases}
\end{equation}
with
\begin{align}
    \Tini &\simeq \Tc \left(\frac{\Tc}{\Tend}\right)^\frac{12 - 8x}{3 + 2x},\\
    \frac{\Rend}{\Rc} &\simeq \left(\frac{\Tc}{\Tend}\right)^\frac{8}{3+2x},\\
    \frac{\Rc}{\Rini} &\simeq \frac{\Tini}{\Tc}\,.
\end{align}
As we are interested in the general effects of such histories on axion production, we will consider a range of dilution factors corresponding to NSC periods of different duration. Although in a realistic history there will be model-dependent limitations on the initial NSC temperature from considerations such as the scale of inflation, we will neglect these in order to characterize the full impact our histories can possibly have on axion production.

Having analytically understood the evolution of the NSC with a time-dependent decay width, we will analyze the resultant dynamics of the axion energy density and the misalignment mechanism in the following section.

\section{Analytical Approach}
\begin{figure}
    \def\sepf{0.61}
	\centering
    \includegraphics[scale=\sepf, trim = 0.2cm 0cm 0.8cm 0cm, clip = true]{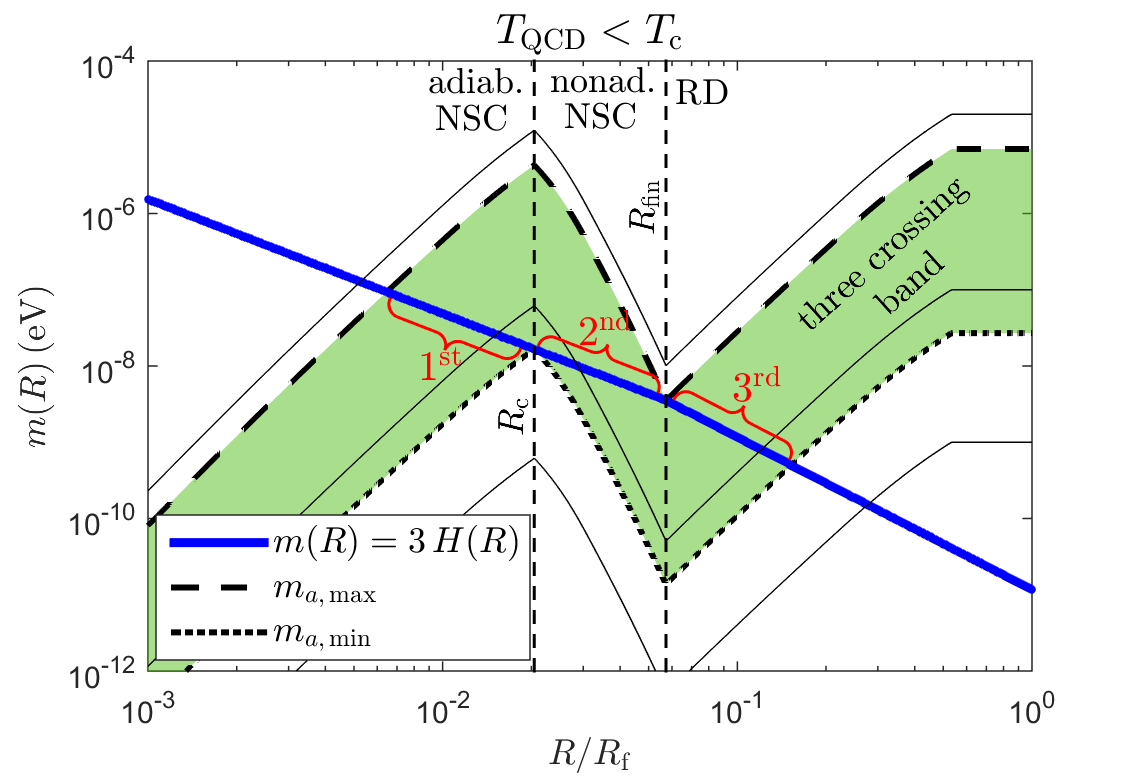}
    \includegraphics[scale=\sepf, trim = 0.2cm 0cm 0.8cm 0cm, clip = true]{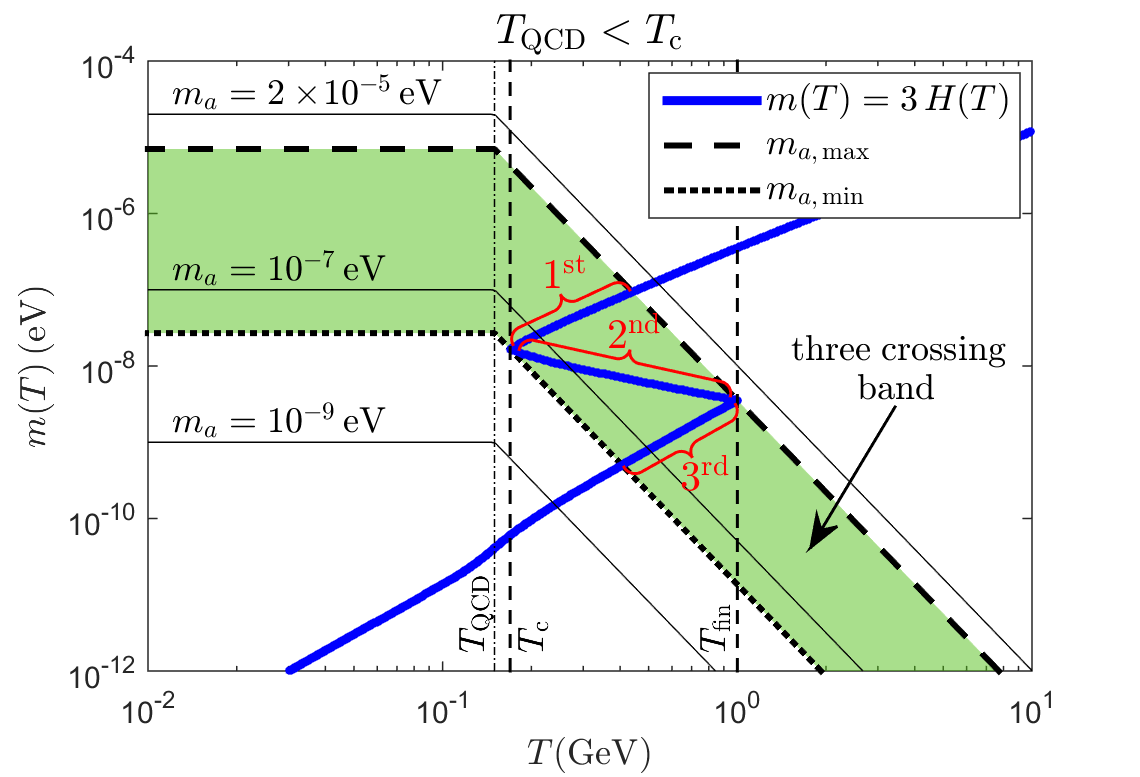}
    \caption{Evolution of the QCD axion mass (black lines) as a function of the scale factor (left) and the SM temperature (right), for $x = -23/2$, $\Tc = 170$~MeV, and $\Tend = 1$~GeV.
    This is an example of the case where $\Tqcd < \Tc$.
    The thick blue lines correspond to $3\, H$.
    The green shaded bands show the values of $m_a$ that result in three crossings of $3\, H(R) = m(R)$.
    The three thin solid black curves show one ($\ma = 10^{-9}$~eV and $\ma = 2 \times 10^{-5}$~eV) or three ($\ma = 10^{-7}$~eV) crossings.
    }
	\label{fig:mass}
\end{figure}

\begin{figure}
    \def\sepf{0.61}
	\centering
    \includegraphics[scale=\sepf, trim = 0.2cm 0cm 0.8cm 0cm, clip = true]{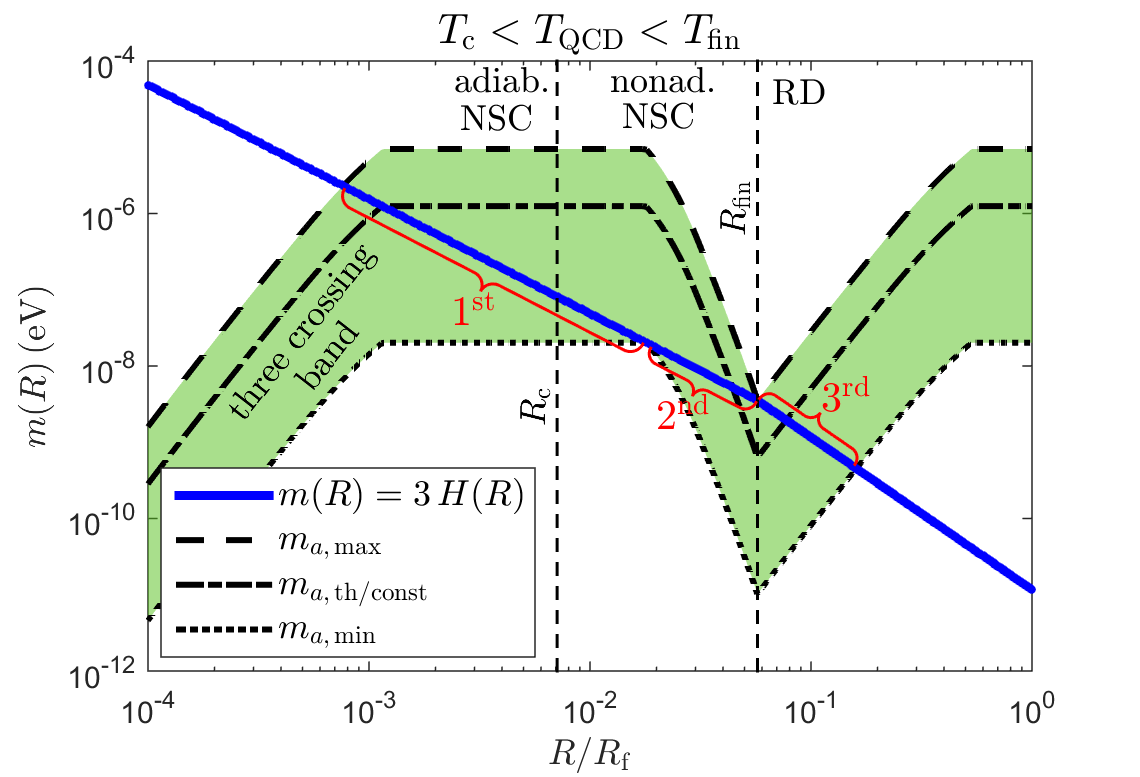}
    \includegraphics[scale=\sepf, trim = 0.2cm 0cm 0.8cm 0cm, clip = true]{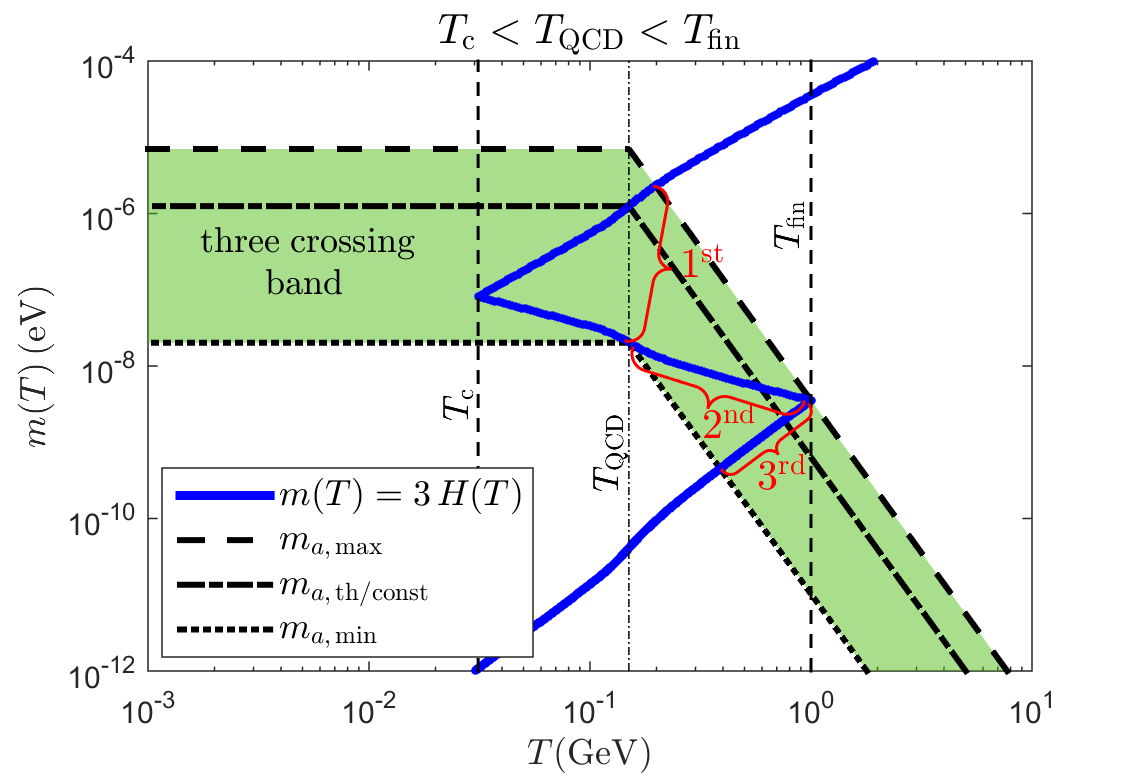}
    \caption{
    Same as Fig.~\ref{fig:mass} but for $\Tc = 31$~MeV.
    This is an example of the case where $\Tc < \Tqcd < \Tend$.
    }
	\label{fig:mass2}
\end{figure}

{\label{sec:analytic}}
Now, we turn to the misalignment mechanism in the context of the NSC described in the previous section. 
Due to its temperature dependence above \(\Tqcd\), the axion mass inherits the modified behavior of the temperature as a function of the scale factor of the Universe. As long as the background temperature undergoes a period of increase, this results in the possibility of multiple crossings of the axion mass and the Hubble expansion rate such that the condition \(3\,H(T) = m(T)\) is satisfied at multiple times, as shown in Figs.~\ref{fig:mass} and~\ref{fig:mass2}.

The evolution of the QCD axion mass (black lines) as a function of the scale factor (left) and the SM temperature (right), for $x = -23/2$, $\Tend = 1$~GeV, and $\Tc = 170$~MeV or $\Tc = 31$~MeV, is shown in Figs.~\ref{fig:mass} and~\ref{fig:mass2}, respectively, as examples of the two cases $\Tqcd < \Tc$ and $\Tc < \Tqcd< \Tend$ (described below).
The three thin black lines in Fig.~\ref{fig:mass} correspond to $\ma = 10^{-9}$~eV, $\ma = 10^{-7}$~eV, and $\ma = 2 \times 10^{-5}$~eV.
When plotted as a function of the temperature, the axion mass shows the expected behavior corresponding to Eq.~\eqref{eq:thermal_mass2}: it grows until $T = \Tqcd$ and then stabilizes to a constant value.
However, it features a nontrivial evolution as a function of the scale factor: $\ma$ can decrease if the SM temperature increases during the evolution of the Universe.
In such a scenario, the mass of the axion can be equal to the Hubble expansion rate $H(R)$ (thick blue lines in the figures) up to three times. 
The green shaded bands show the values of $m_a$ that result in three crossings of $3\, H(R) = m(R)$.
The three thin solid black curves show one ($\ma = 10^{-9}$~eV and $\ma = 2 \times 10^{-5}$~eV) or three ($\ma = 10^{-7}$~eV) crossings. Note that the blue $H(R)$ line experiences a change in slope at \(R = \Rend\) once RD is restored, as seen in Fig.~\ref{fig:bkg2}. Additionally, we take into account the temperature dependence of the relativistic degrees of freedom \(\gs(T)\), which is responsible for the slight bend of $H(T)$ at \(T \sim \Tqcd\) in the right panels. 

Multiple crossings correspond to scenarios in which axion oscillations can pause for an extended time before resuming after the final crossing. 
They are only possible if the background temperature increases during the nonadiabatic phase of the NSC, {\it i.e.} if $x < -3/2$. 
Furthermore, the dependence of $m(R)$ must be steeper than $H(R)$ during the nonadiabatic phase, resulting in the condition $x < -3$. We note that if the equality is realized, the cosmology allows for an extended period of $3\, H \approx m$ rather than multiple discrete crossings.
From a physical point of view, if the condition $m < 3\, H$ is satisfied for a second time during the cosmological history of the axion, it will cause a stop in oscillations. If at this point the field has no kinetic energy left, such that $\dot \theta\sim 0$, the amplitude will freeze into a new constant value. On the other extreme, if there is only kinetic energy stored in the field, the amplitude can vary during this period, and the energy density will redshift as $R^{-6}$~\cite{Co:2019jts, Chang:2019tvx, Barman:2021rdr}. Furthermore, once $3\, H = m$ is met again, oscillation will resume from a -- in general -- different ``initial'' state than in the first crossing, where this new state depends on the energy configuration of the field at the moment it became stuck for a second time. We will expand upon this discussion in the subsequent sections. 

In what follows, for fixed NSCs, we will categorize possible scenarios by the number of crossings -- three, two, or one -- as well as by the minimum temperature reached during the NSC, \(\Tc\), and the temperature at the end of NSC, $\Tend$, as compared to \(\Tqcd\).

\subsection{Three Crossings}
In this subsection, we will define the conditions that allow for there to be three crossings, and we will describe the properties of the resultant scenarios. 
The range of parameter values that corresponds to three crossings can be determined for a given history. As seen in Figs.~\ref{fig:mass} and~\ref{fig:mass2}, there are two general configurations of the axion mass that can support three crossings, depending on the values of \(\Tc\) and \(\Tend\). These are: \(\Tqcd < \Tc\) in which the mass depends on the temperature throughout the NSC, and \(\Tc < \Tqcd < \Tend\) in which the mass of the axion becomes constant for some time during the NSC due to the temperature temporarily dropping below \(\Tqcd\). For parameter values in which three crossings do occur, these two cases are primarily distinguished by the behavior of the first crossing, which can occur during the temperature-dependent or constant regimes of the axion mass and during the adiabatic or nonadiabatic NSC phases. The second and third crossings, however, always occur in the nonadiabatic phase of NSC and in RD after NSC, respectively, both in the regime of temperature-dependent axion mass. 

\subsubsection*{\boldmath Case $\Tqcd < \Tc$}
In this case, the axion mass depends on temperature throughout the entire NSC.
As seen in Fig.~\ref{fig:mass}, the first crossing always occurs in the adiabatic NSC phase with a temperature-dependent axion mass (which is trivially satisfied because all temperatures of interest are above \(\Tqcd\) in this case). However, the first crossing cannot occur in the RD phase prior to NSC because \(\Tini > \Tend\) even if the temperature increases during the nonadiabatic phase. 

For a given thermal history, the extent of the value of $m_a$ corresponding to three crossings can be easily defined by two limiting cases.
The smallest allowed value of $m_a$ can be found by setting \(3\, H = m\) at \(T = \Tc\), and is given by
\begin{equation} \label{eq:ma_min1}
    m_{a,\,{\rm min}} \approx 3 H_r(\Tend) \left(\frac{\Tc}{\Tend}\right)^\frac{12}{3 + 2x} \left(\frac{\Tc}{\Tqcd}\right)^4 .
\end{equation}
Any lower value of $m_a$ will result in a total of one crossing that occurs in RD after NSC, thus reproducing the standard scenario described in Section~\ref{sec:standard}.
On the contrary, the largest value of $m_a$ is similarly obtained by setting \(3\, H = m\) at \(T = \Tend\): 
\begin{equation} \label{eq:ma_max}
    m_{a,\,{\rm max}} \approx 3 H_r(\Tend) \left(\frac{\Tend}{\Tqcd}\right)^4,
\end{equation}
where any larger value will result in one single crossing, which occurs in the adiabatic phase of the NSC with a temperature-dependent axion mass. These two limiting values are shown in Fig.~\ref{fig:mass} and define the boundaries of the green three-crossing band. 

\subsubsection*{\boldmath Case $\Tc < \Tqcd < \Tend$}
In this case, as seen in Fig.~\ref{fig:mass2}, the first crossing can occur in the adiabatic phase with either a temperature-dependent or constant axion mass, or in the nonadiabatic phase with a constant mass. As before, the first crossing cannot occur prior to the NSC because of the temperature hierarchy. 

The smallest value of $m_a$ that allows for three crossings is now determined by the transition between the temperature-dependent and constant mass regimes within the nonadiabatic phase, rather than \(\Tc\), and is given by: 
\begin{equation} \label{eq:ma_min2}
    m_{a,\,{\rm min}} \approx 3 H_r(\Tend) \left(\frac{\Tqcd}{\Tend}\right)^\frac{12}{3 + 2x}.
\end{equation}
This again corresponds to the boundary between three crossings and only one in RD after NSC, with lower values of $m_a$ reproducing the standard RD misalignment. 

The highest three-crossing value of $m_a$ in this case is given by Eq.~\eqref{eq:ma_max} as before because the second and third crossings always occur in the temperature-dependent mass regime, as \(\Tend > \Tqcd\). However, this maximum value does not necessarily mark the transition to one crossing in adiabatic NSC with a temperature-dependent mass due to the presence of the constant mass regime and the possible configurations of the first crossing. The transition between the temperature-dependent and constant-mass regimes of the adiabatic phase is given by:
\begin{equation} \label{eq:ma_therm_const}
    m_{a,\,{\rm th/const}} \approx 3 H_r(\Tend) \left(\frac{\Tc}{\Tend}\right)^\frac{12}{3 + 2x} \left(\frac{\Tqcd}{\Tc}\right)^{3/2}.
\end{equation}
If this value is smaller than \(m_{a,\,{\rm max}}\), then the range of $m_a$ corresponding to three crossings is split into two regions based on the relevant axion mass regime for the first crossing: constant mass for \(m_{a,\,{\rm min}} < m_a < m_{a,\,{\rm th/const}}\), and temperature dependent for \(m_{a,\,{\rm th/const}} < m_a < m_{a,\,{\rm max}}\). In this case, the maximum mass therefore corresponds to the boundary between three crossings and only one crossing in the adiabatic phase with a temperature-dependent mass, as was the case with \(\Tc > \Tqcd\). If instead \(m_{a,\,{\rm th/const}} > m_{a,\,{\rm max}}\), then the maximum mass marks the boundary between three crossings and only one crossing in the NSC (either adiabatic or nonadiabatic phase) with a constant mass, as the transition of the mass regimes occurs outside of the three-crossing range. It is worth noting that the transition between adiabatic and nonadiabatic NSC with a constant axion mass, corresponding to \(m_a \approx 3\, H(\Tc) \equiv 3\, H_{\rm c}\), will affect some details but in general is not very significant. 

\subsubsection*{Oscillation Temperature}
As previously mentioned, if three crossings occur, the second and third crossing are always in the nonadiabatic phase of NSC and the subsequent RD phase, respectively, both with temperature-dependent axion mass. The first crossing, however, can occur in a variety of cases. We denote the temperature of the Universe at the time of a crossing by \(T_i\) with \(i = \) 1, 2, or 3, which is implicitly defined by the equality 
\begin{equation}
    3\, H(T_i) = m(T_i)\,,
\end{equation}
and we correspondingly denote the scale factor at the time of each crossing as \(R_i\). Using Eqs.~\eqref{eq:thermal_mass2}, \eqref{eq:Hubble}, and~\eqref{T(a)} we obtain the following expressions for the temperature at the time of each crossing:
\begin{equation} \label{eq:T1}
    T_1 \simeq 
    \begin{dcases}
        \left[\frac{m_a}{3\, H_r(\Tend)} \left(\frac{\Tend}{\Tc}\right)^\frac{12}{3 + 2x} \left(\frac{\Tqcd}{\Tc}\right)^4\right]^{2/11} \Tc & \text{ for } R_1 < \Rc \text{ and } T_1 > \Tqcd\,,\\
        \left(\frac{m_a}{3\, H_r(\Tend)}\right)^\frac23 \left(\frac{\Tend}{\Tc}\right)^\frac{8}{3 + 2x} \Tc & \text{ for } R_1 < \Rc \text{ and } \Tqcd > T_1\,,\\
        \left(\frac{m_a}{3\,H_r(\Tend)}\right)^\frac{3 + 2x}{12} \Tend & \text{ for } \Rc < R_1\,.
    \end{dcases}
\end{equation}
The first case happens when the mass has a temperature dependence ($T_1 > \Tqcd$), in the adiabatic phase ($R_1 < \Rc$).
However, the second case also corresponds to the adiabatic phase but occurs when the mass is already constant ($T_1 < \Tqcd$).
Finally, the last possibility corresponds to a crossing in the nonadiabatic phase ($R_1 > \Rc$) with a constant mass.
Having three crossings also requires $R_1 < \Rend$.
We note that a first crossing in the nonadiabatic phase with a time-dependent mass can not occur because that would require a mass \(\ma\) smaller than the three-crossing minimum value (Eqs.~\eqref{eq:ma_min1} and~\eqref{eq:ma_min2}).
The second and third crossings are given by
\begin{align} 
    T_2 &\simeq \Tend \left[\frac{m_a}{3\,H_r(\Tend)} \left(\frac{\Tqcd}{\Tend}\right)^4\right]^\frac{3 + 2x}{4\, (6 + 2x)}, \label{eq:T2}\\
    T_3 &\simeq \Tend \left[\frac{m_a}{3\, H_r(\Tend)} \left(\frac{\Tqcd}{\Tend}\right)^4\right]^\frac16. \label{eq:T3}
\end{align}
We notice that the second and third crossings always occur between $\Tqcd$ and $\Tend$.
We further note that in all cases, the temperatures of the three crossings maintain the hierarchy $T_1 < T_2 < T_3$.
Finally, as the third crossing occurs in a RD era, Eq.~\eqref{eq:T3} coincides with Eq.~\eqref{eq:Tosc_std}, in the case $\Tosc \gg \Tqcd$.

\subsubsection*{Relic Abundance}
With the conditions for there to be three crossings defined, we will now determine the current abundance of axions by considering the evolution of the energy density. Recall that we define the scale factor at the time of the first, second, and third crossings by \(R_i\) with \(i =\) 1, 2, and 3 respectively, which is determined by the condition \(3\, H(R_i) = m(R_i)\). At the time of the first crossing, just before the onset of the first phase of oscillations, the axion energy density is given by
\begin{equation}
    \rho_a(R_1) = m(R_1)\, n(R_1) \simeq \frac12 f_a^2\, \theta_{\rm i}^2\, m^2(R_1)\,.
\end{equation}
In the period between the first and second crossings, the axion field oscillates and its number is conserved. The energy density at the time of the second crossing is therefore obtained simply from redshift as 
\begin{equation}
    \rho_a(R_2) = m(R_2)\, n(R_2) = m(R_2)\, n(R_1)\, \frac{s(R_2)}{s(R_1)}\, \frac{S(R_1)}{S(R_2)} = \rho_a(R_1)\, \frac{m(R_2)}{m(R_1)}\, \frac{s(R_2)}{s(R_1)}\, \frac{S(R_1)}{S(R_2)}\,.
\end{equation}
In the period between the second and third crossings, the axion field is frozen again and oscillations are temporarily stopped. At the beginning of this period, when \(R = R_2\), the axion may have developed a significant kinetic energy depending on the phase of the previous oscillations. The energy density at \(R = R_3\) is therefore given by a mixture of kinetic and potential energy density, which evolve differently, with potential remaining constant up to changes in the axion mass while kinetic scales as \(R^{-6}\). We parameterize this by introducing \(0 \leq \alpha \leq 1\) as the fraction of the total energy density in the form of potential energy at \(R = R_2\). Taking this into account, the energy density at the third crossing is given by 
\begin{equation}
    \rho_a(R_3) = \alpha\, \rho_a(R_2) \left[\frac{m(R_3)}{m(R_2)}\right]^2 + (1-\alpha)\, \rho_a(R_2) \left[\frac{s(R_3)}{s(R_2)} \frac{S(R_2)}{S(R_3)}\right]^2,
\end{equation}
where the first term is the fraction of potential energy density, while the second term is kinetic. 
Note that $\alpha$ {\it is not } a free parameter, but depends on the details of the oscillations at the second crossing, namely \(\theta(R_2)\). 
Finally, after \(R = R_3\), the axion field resumes its oscillations and the present energy density is given by 
\begin{equation}
    \rho_a(R_0) = \rho_a(R_3)\, \frac{m_a}{m(R_3)}\, \frac{s(R_0)}{s(R_3)}\, \frac{S(R_3)}{S(R_0)}\,.
\end{equation}
Combining the previous expressions, the axion energy density at present becomes
\begin{align}\label{eq:rho_full}
    \rho_a(R_0) &\simeq \frac12 f_a^2\, \theta_{\rm i}^2\, m_a\, m(R_1)\, \frac{s(R_0)}{s(R_1)} \times \frac{S(R_1)}{S(\Rend)} \nonumber \\
    &\qquad\times \frac{m(R_2)}{m(R_3)}\, \frac{s(R_2)}{s(R_3)}\, \frac{S(R_3)}{S(R_2)} \left[\alpha\, \left[\frac{m(R_3)}{m(R_2)}\right]^2 + (1-\alpha)\, \left[\frac{s(R_3)}{s(R_2)} \frac{S(R_2)}{S(R_3)}\right]^2\right].
\end{align}
The first factor corresponds to the standard case that does not have an injection of entropy, the second to the production of entropy due to the nonadiabatic phase, and the last one appears only in the case where $3\, H = m$ is realized multiple times.
Interestingly, the latter factor can enhance or suppress the energy density for potential or kinetic energy domination, respectively.

\subsection{Two Crossings}
As we have seen above, the conditions given in Eqs.~\eqref{eq:ma_min1} to~\eqref{eq:ma_min2} separate scenarios with three crossings from those with only one single crossing. These are therefore special cases, which feature only two crossings, where the first and second, or the second and third, crossings occur simultaneously, while the other remains distinct. 

The case where \(R_2 = R_3 = \Rend\) reproduces Eq.~\eqref{eq:ma_max} and gives a present energy density of 
\begin{equation}
    \rho_a(R_0) \simeq \frac12 f_a^2\, \theta_{\rm i}^2\, m_a\, m(R_1)\, \frac{s(R_0)}{s(R_1)} \times \frac{S(R_1)}{S(\Rend)} \,.
\end{equation}
In general, the first crossing can occur in either the adiabatic phase of NSC (with temperature-dependent or constant axion mass) or the nonadiabatic phase (with constant mass only). When \(R_1 \leq R_{\rm c}\), we see that the expression above corresponds to the standard case with {\it maximal} entropy dilution, whereas when \(R_1 > R_{\rm c}\) the dilution factor is smaller. Note that the axion does not enter a prolonged period of paused oscillations, as \(3\, H = m\) occurs for the second and third time simultaneously. 

Alternatively, when the first and second crossings occur simultaneously, there are two possible cases which coincide with the two cases of the previous subsection. If \(\Tqcd < T_{\rm c}\), then \(R_1 = R_2 = R_{\rm c}\) reproduces Eq.~\eqref{eq:ma_min1}, while if \(T_{\rm c} < \Tqcd < \Tend\), then \(R_1 = R_2\) occurs at \(T = \Tqcd\) in the nonadiabatic phase, reproducing Eq.~\eqref{eq:ma_min2}. In both cases, the present axion energy density is  
\begin{equation}
    \rho_a(R_0) \simeq \frac12 f_a^2\, \theta_{\rm i}^2\, m_a\, m(R_3)\, \frac{s(R_0)}{s(R_3)}
\end{equation}
which corresponds to the standard case of a RD cosmological scenario. We have taken the limit \(\alpha \to 1\) because axion oscillations are essentially delayed until \(R = R_3\), and therefore the axion does not develop a significant kinetic energy at \(R = R_1 = R_2\).

\subsection{One Crossing}
The case of only one crossing is essentially unchanged from the standard case apart from small details, and therefore we discuss it only briefly. If \(3\, H = m\) occurs in RD after the NSC phase, then the situation is the same as the standard case presented in Section~\ref{sec:standard}. If \(3\, H = m\) instead happens before NSC or during adiabatic NSC, then the situation is basically the same as the standard NSC picture ({\it i.e.} with \(x = 0\)) with maximal dilution (though the oscillation temperature can be different if the crossing occurs at temperatures below \(\Tc\) of the \(x = 0\) case). The main difference between these standard scenarios and the case of a time-dependent decay width is the effect on the nonadiabatic phase. 

There is one case where a single crossing can occur during the nonadiabatic NSC phase, and it is only possible if \(T_{\rm c} < \Tqcd < \Tend\) and \(3\, H_{\rm c} > m_{a,\,{\rm max}}\) (see the discussion below Eq.~\eqref{eq:ma_therm_const}). In this case, \(3\, H = m\) is satisfied in the constant mass regime. Otherwise, if three crossings are allowed by the cosmological history, a scenario with a total of only one crossing occurring in the nonadiabatic phase of NSC is not possible. In passing, we note that it is generally possible to have only one crossing during nonadiabatic NSC even if the temperature increases (or remains constant), but this corresponds to the histories with $x > -3$, which have not been our focus, as they do not support multiple instances of \(3\, H = m\).

Lastly, if \(\Tend < \Tqcd\), then the entire nonadiabatic phase occurs at temperatures below \(\Tqcd\) such that the mass remains unaffected. In this case it is only possible to have a total of one crossing, which can occur anywhere in the history. We will come back to this case toward the end of the next section. 

\section{Numerical Results} \label{sec:results}
Up until this point we have been discussing the axion misalignment mechanism in our NSCs analytically. In this section, we present numerical solutions of the axion equation of motion in the background of the NSC described in Section~\ref{sec:NSC}. We simultaneously solve Eqs.~\eqref{eq:axion_eom}, \eqref{eq:boltzmann1}, and~\eqref{eq:boltzmann2}, approximating the thermal axion mass by Eq.~\eqref{eq:thermal_mass2} and taking into account that the axion energy density remains small compared to the background. An example of the background evolution is shown in Figs.~\ref{fig:bkg_rho} and~\ref{fig:bkg2}. 

\begin{figure}
    \def\sepf{0.9}
	\centering
    \includegraphics[scale=\sepf, trim = 0.5cm 0cm 0.5cm 0.5cm, clip=true]{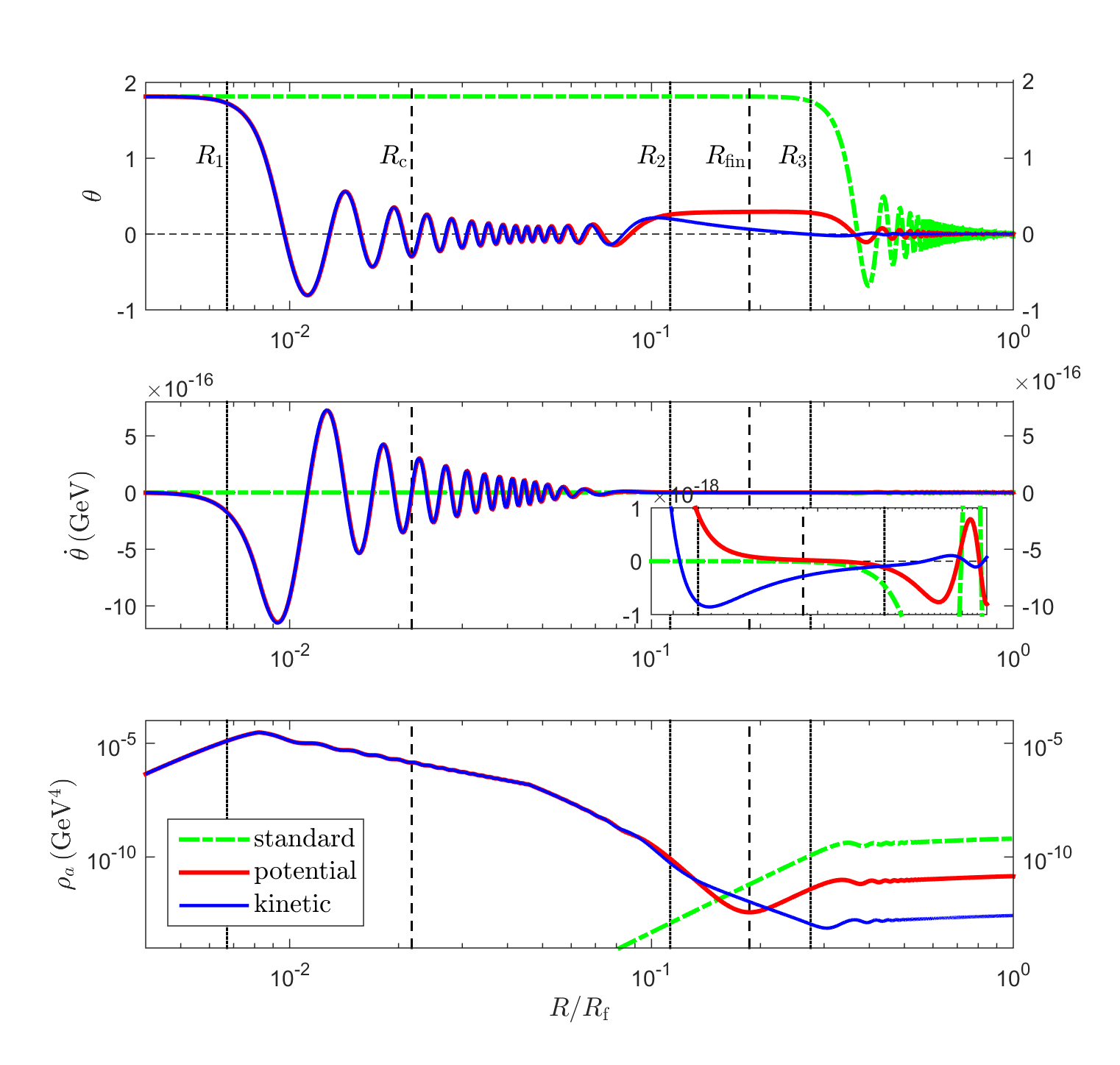}
    \caption{
    Evolution of the misalignment angle $\theta$, its velocity $\dot \theta$, and the axion energy density $\rho_a$ as a function of the scale factor $R$, for two different three-crossing benchmarks (shown in solid blue and red) as well as the corresponding case for a standard RD cosmology (dashed green). 
    All curves have \(\theta_{\rm i} = \pi/\sqrt{3}\). Solid curves additionally have \(x = -23/2\), \(T_{\rm fin} \approx 1\)~GeV, and \(\Tc \approx 87\)~MeV, but the mass is varied slightly: \(m_a = 1.10 \times 10^{-6}\)~eV (blue) and \(m_a = 1.09 \times 10^{-6}\)~eV (red). 
    The solid curves show the dominant kinetic and potential cases for the energy density between the second and third crossings, respectively. 
    Notice that the final energy density for both of the solid curves lies below the standard-cosmology value, indicating that a smaller axion mass would be required in order to obtain the same energy density. 
    } 
    \label{fig:individual_sample4}
\end{figure}
In Fig.~\ref{fig:individual_sample4} below, we show the evolution of the axion misalignment angle \(\theta\) (upper panel), its velocity \(\dot\theta\) (central panel), and the axion energy density \(\rho_a\) (lower panel) as functions of the scale factor $R$ for two sets of example parameters that lead to three crossings of the Hubble rate and the mass of the axion. 
The two solid curves shown in each panel have the same values of the following parameters: \(\theta_{\rm i} = \pi/\sqrt{3}\), \(x = -23/2\), \(\Tc \approx 87\)~MeV, and \(T_{\rm fin} \approx 1\)~GeV; however they differ slightly in the value of the zero-temperature axion mass as follows: \(m_a = 1.10 \times 10^{-6}\)~eV (blue) and \(m_a = 1.09 \times 10^{-6}\)~eV (red).\footnote{We note that the particular value of \(\theta_{\rm i}\) chosen here is unimportant, provided that it is the same for all curves shown, therefore we have neglected any effects due to anharmonicities in the axion potential for large \(\theta_{\rm i}\).} For comparison, in dashed green, we show the corresponding standard history case, with the same values of \(\theta_{\rm i}\) and \(\ma\) as above. Notice that for such slight variation of \(\ma\), no change is visible for the standard case, while the two solid curves display significant differences.
Additionally, the three dotted vertical lines correspond to $R = R_1$, $R = R_2$, and $R = R_3$, whereas the dashed lines to $R = \Rc$ and $R = \Rend$. Finally, recall that we normalize the scale factor to an arbitrary value \(R = R_{\rm f}\) occurring well after the end of NSC, such that all thermal histories coincide once standard RD is established.

Following along the evolution in the figure, both of the two solid curves begin to oscillate shortly after the time of the first crossing
(shown by the vertical black dotted line marked by \(R_1\)).
Oscillations occur during the period \(R_1 < R < R_2\) when the Hubble rate is smaller than the axion mass. However, due to the increasing temperature of the nonadiabatic NSC phase, the axion mass decreases correspondingly for some time such that it falls below the Hubble rate at the time of the second crossing (shown by the vertical black dashed line marked \(R_2\)). Oscillations after this point are temporarily paused due to the restoration of Hubble friction, and they remain so until the axion mass once again increases after the end of the NSC phase. Once the axion mass returns to being larger than the Hubble rate after the third and final crossing (shown by the vertical black dashed line marked \(R_3\)), oscillations resume and proceed as in the standard history, albeit with a different amplitude.
We note that for the two benchmarks, the precise timing of the crossings is slightly different because of the difference in mass; however, this is too small to be noticeable in the figure. 

As seen in the figure, the two solid curves exhibit different behaviors during the intermediate period between the second and third crossings, \(R_2 < R < R_3\), when oscillations are temporarily paused.
As mentioned in the previous section, these behaviors are distinguished by the dominant form of the axion energy density during this period, and we parameterize this dependence by defining the parameter \(\alpha\) as the fraction of the axion energy density that is in the form of potential energy at the time of the second crossing, when oscillations begin to stop. 
This can be explicitly seen in the two curves as follows. 

During the intermediate period, the angle of the blue curve is seen to decrease with a corresponding nonzero velocity, whereas the angle of the red curve stays roughly constant with a largely unchanging velocity near zero. Furthermore, the energy density of the blue curve decreases with a constant slope of \(\rho_a \propto R^{-6}\) throughout this period whereas the energy density of the red curve follows the change in mass due to the temperature, initially decreasing when the temperature is increasing (with a slope approaching \(\rho_a \propto R^{-20}\), though this is not reached in the figure due to the slow curvature of the numerical \(m(T)\) curve during this period) followed by an increase as \(\rho_a \propto R^8\) after the end of the NSC. 
The solid blue curve therefore corresponds to dominant kinetic energy with \(\alpha \approx 0\), and the solid red curve to dominant potential energy with \(\alpha \approx 1\). 
The two curves serve as examples of these two behaviors, however, it is possible to have cases that transition between kinetic- and potential-dominated regimes. Such mixed cases fill out the space in-between the two solid curves, and in some cases can actually extend to energy densities that sit slightly higher than the final potential value, but still well below the standard curve. 
In all cases, particularly in the kinetic case, the three-crossing history dynamically drives the angle toward a smaller value that acts as a new ``initial'' condition for the final phase of oscillations. Furthermore, the velocity tends toward zero during the intermediate period, so that it is near zero at the third crossing. These effects combine to essentially reset the misalignment mechanism at the time of the third crossing such that the final oscillations begin with a value of the angle that is smaller than it was originally. 

We note that in the bottom panel of Fig.~\ref{fig:individual_sample4} the final energy densities of all curves straighten out to a constant slope of \(\rho_a \propto R\) rather than \(\rho_a \propto R^{-3}\) because the temperature has not yet dropped below \(\Tqcd\) for the final time, and the axion mass is therefore still changing. Once the mass becomes constant again, the energy densities follow the matter-like redshift relation as expected. 
The final axion abundance is determined by the value of \(\theta\) at the time of the third crossing when oscillations resume, or equivalently by the maximum amplitude of oscillations after the final crossing. However, the dependence of this abundance on \(\alpha\) is not trivial due to a strong sensitivity to the details of the precise phase of oscillations. Slight changes to the input parameters, as seen by the slightly different values of \(\ma\) of the two solid curves of Fig.~\ref{fig:individual_sample4}, can significantly alter the behavior of the intermediate period, resulting in a different abundance by potentially multiple orders of magnitude once oscillations resume. 
However, the final abundance in our increasing-temperature NSC scenarios is always smaller than the corresponding abundance in a standard RD history (for the same initial angle and zero-temperature mass). Therefore, due to the inverse relationship between the axion abundance and its mass, in order for our NSC scenarios to reach the same final abundance as in the standard case, we see that the zero-temperature axion mass $m_a$ must always be lower than the standard value. Thus, if we require that our scenarios lead to the correct abundance of DM in our Universe today, the resultant axion mass will be smaller than (as well as overlapping with) the standard-history window, as expected because of entropy dilution. 

\begin{figure}
    \def\sepf{0.99}
	\centering
    \includegraphics[scale=\sepf]{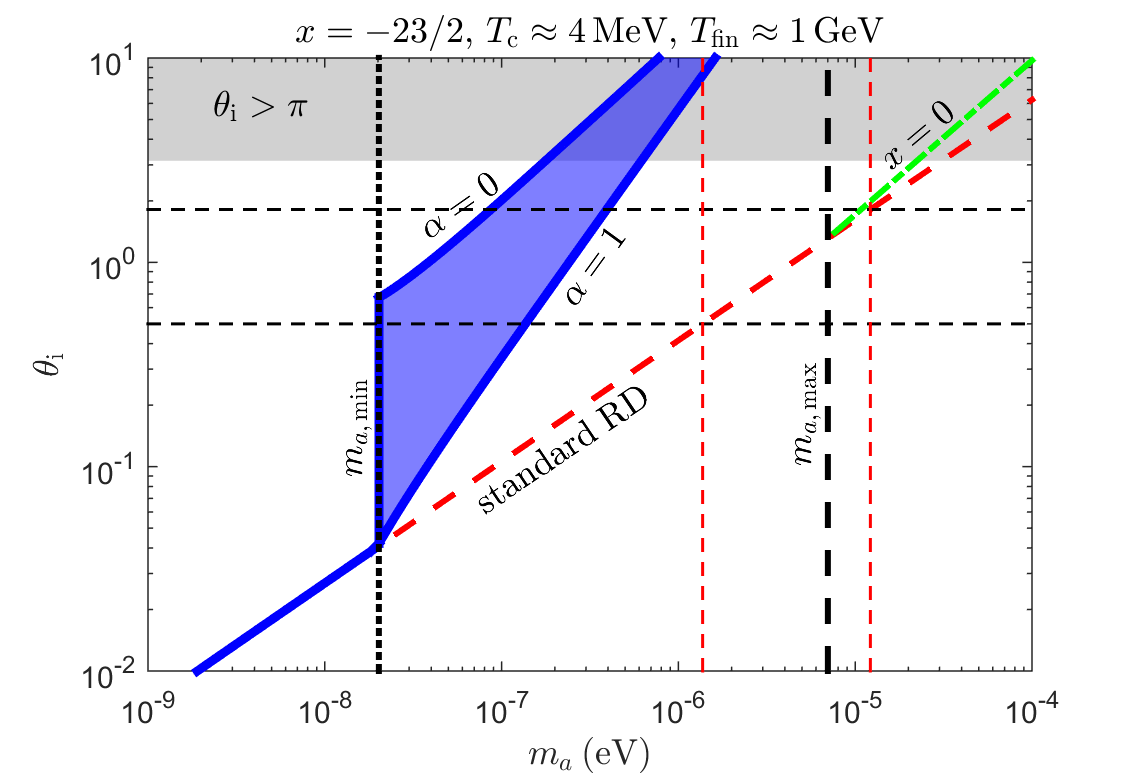}
    \caption{Correct DM abundance in the \(\theta_{\rm i}\) -- $m_a$ plane for an example increasing-temperature NSC (solid blue), with corresponding \(x = 0\) NSC (dash-dot green), and standard RD (dashed red) shown as well. Three-crossings occur between \(m_{a{\rm,min}}\) and \(m_{a{\rm,max}}\), shown by the vertical black lines. The two thin red dashed lines show the standard range of \(m_a\) for the range of initial angle \(1/2 \leq \theta_{\rm i} \leq \pi/\sqrt{3}\) (thin dashed black lines). All lines are analytical with \(\gs(T)\) taken into account. 
    }
    \label{fig:theta_i_m_a_2}
\end{figure}

With this expectation, we will now explore the parameter space that reproduces the entire current DM abundance by requiring \(\Omega_a\, h^2 = 0.12\)~\cite{Planck:2018vyg}. In Fig.~\ref{fig:theta_i_m_a_2} we show an analytical band (blue) in the \(\theta_{\rm i}\) -- $m_a$ plane that reproduces the observed DM abundance for a history that supports three crossings, with \(x = -23/2\), \(\Tc \approx 4\)~MeV, and \(\Tend \approx 1\)~GeV.
The blue shaded region corresponds to values of the mass that lead to three crossings with \(m_{a{\rm, min}} \leq m_a \leq m_{a{\rm, max}}\), where the maximum and minimum masses are given by Eqs.~\eqref{eq:ma_max} and~\eqref{eq:ma_min2}, respectively. For \(m_a < m_{a{\rm,min}}\), the blue region becomes a line and merges with the standard history shown by the diagonal red dashed line. Note that where the blue band intersects with the range \(1/2 \leq \theta_{\rm i} \leq \pi/\sqrt{3}\), the values of $m_a$ that are needed to obtain the correct DM abundance are smaller than the standard range (shown by the vertical red dashed lines) by one to two orders of magnitude.\footnote{For clarity, we note that even for \(\theta_{\rm i} = \pi/\sqrt{3}\), which corresponds to the post-inflationary PQ breaking scenario, the axion mass is still smaller than that of the standard-history window.} We also show the corresponding line (dash-dot green) for the correct DM abundance from a NSC with \(x = 0\), where the initial and final NSC temperatures are the same as the \(x = -23/2\) case. This line deviates from the standard history for \(m_a > m_{a{\rm,max}}\) and sits at lower masses than the standard RD history, but at higher masses than the increasing-temperature NSC. 
All lines shown in this figure are obtained from the analytical expressions presented in previous sections, with the temperature dependence of the degrees of freedom \(\gs(T)\) taken into account by iterative convergence using data from Ref.~\cite{Borsanyi:2016ksw}, without including anharmonic effects. We have checked that the full numerical evolution matches the extent of the analytical blue band in the range \(1/2 \leq \theta_{\rm i} \leq \pi/\sqrt{3}\) quite well. 

\begin{figure}
    \def\sepf{0.9}
	\centering
    \includegraphics[scale=\sepf]{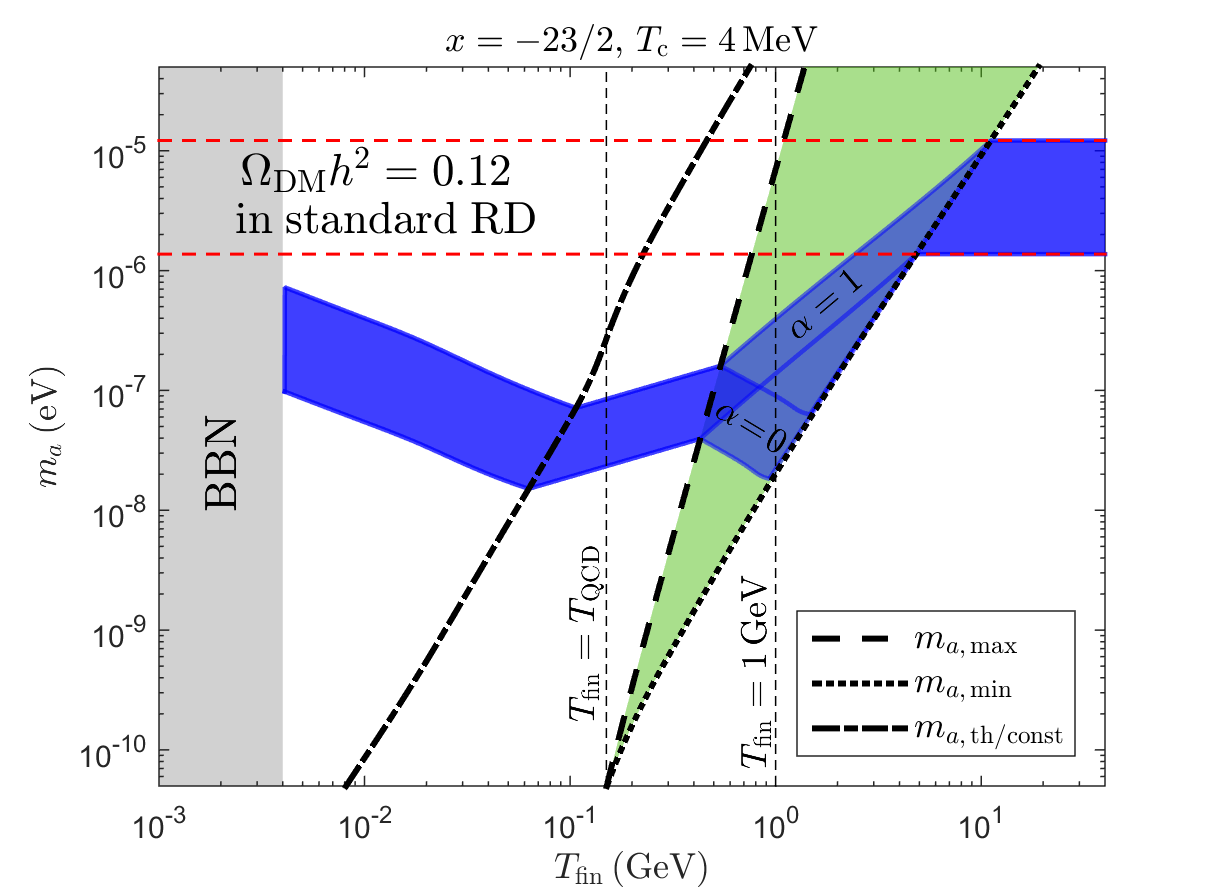}
    \includegraphics[scale=\sepf]{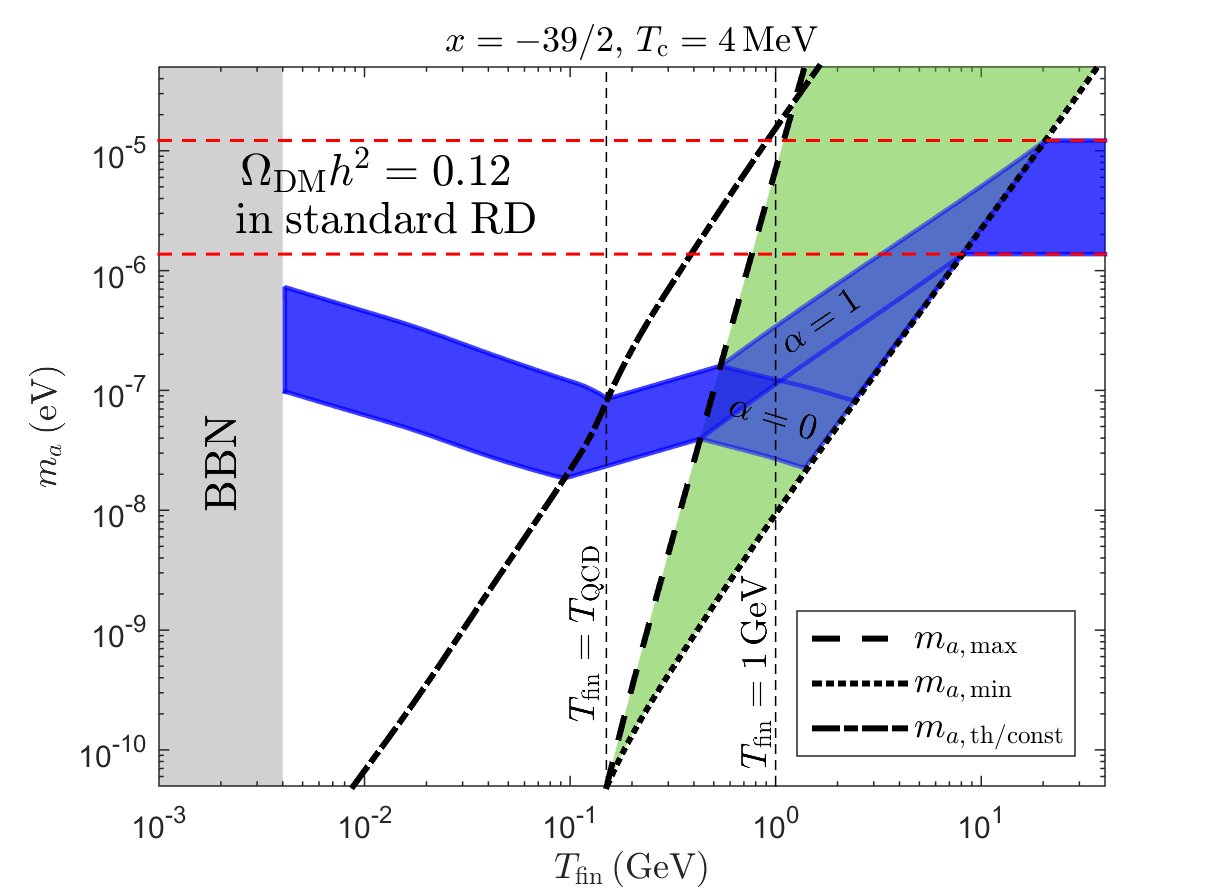}
    \caption{Axion mass vs final NSC temperature. Values within the blue band reproduce the correct DM relic abundance in our increasing-temperature NSC scenarios. The width of the blue band corresponds to the range \(1/2 \leq \theta_{\rm i} \leq \pi/\sqrt{3}\). Scenarios that experience three crossings occur within the green triangular region, while those with only one crossing lie in the white space on either side (one crossing during NSC on the left, one crossing in RD after NSC on the right). 
    } 
    \label{fig:m_Tc}
\end{figure}

To explore the full range of masses accessible to our increasing-temperature NSC scenarios, in Fig.~\ref{fig:m_Tc} we show the correct DM abundance (blue bands) in the $m_a$ -- \(\Tend\) plane for fixed \(\Tc = 4\)~MeV with \(x = -23/2\) (upper panel) and \(x = -39/2\) (lower panel). We have fixed \(\Tc\) to its minimum value as this maximizes the deviation from a standard cosmological history. Note that as \(\Tend\) changes for fixed \(\Tc\) and \(x\), the initial NSC temperature \(\Tini\) changes as well. The thickness of the blue bands is set by the range of initial misalignment angle \(1/2 \leq \theta_{\rm i} \leq \pi/\sqrt{3}\). The three dividing conditions, given by Eqs.~\eqref{eq:ma_max} -- \eqref{eq:ma_therm_const}, are shown by three thick black lines denoted in the legend, and the region that corresponds to three crossings is shaded green. The intersection of the blue band with the green-shaded region shows the parameters for which the correct DM abundance is obtained with three crossings. The region to the right of the green-shaded region corresponds to one single crossing occurring after NSC, thus the blue band aligns with the standard RD window in this region. The area to the left of the green-shaded region corresponds to one single crossing occurring during NSC. Furthermore, the region to the left of the \(m_{a{\rm,th/const}}\) line corresponds to the temperature-dependent regime of the axion mass, while the region to the right corresponds to the constant regime. Note that for \(\Tend < \Tqcd\) the entire nonadiabatic phase occurs at temperatures below \(\Tqcd\), therefore it is not possible to have three crossings for these temperatures. Though the two panels differ in the precise slopes and positions of the various lines and regions, the two values of \(x\) shown do not result in substantial differences.

From Figs.~\ref{fig:theta_i_m_a_2} and~\ref{fig:m_Tc} we can see that the smallest masses that can account for the full DM abundance in a history that experienced three crossings correspond to the \(\alpha \approx 0\) kinetic case discussed above. This makes sense because these scenarios have the smallest values of the misalignment angle at the third crossing and thus the smallest energy densities, for a fixed mass, as compared to the standard history. We can estimate the minimum possible axion mass allowed in our three-crossing scenarios using Eqs.~\eqref{eq:ma_min2} and \eqref{eq:rho_full} in the constant mass regime for \(\Tc < \Tqcd < \Tend\) with \(\alpha = 0\) (noting that the minimum three-crossing mass in Fig.~\ref{fig:m_Tc} is reached where the bottom of the \(\alpha = 0\) band meets \(m_{a{\rm,min}}\) near \(\Tend = 1\)~GeV):
\begin{equation} \label{eq:ma_minimum}
    m_a \gtrsim \frac{ \Tqcd^2}{M_{\rm P}} \left(\frac{45\, (f_a\, m_a)^2\, \theta_{\rm i}^2\, s_0}{4\pi^2\, \Omega_{\rm DM}\, \rho_{\rm crit}\, \Tqcd^3}\right)^{\frac{6\, (2x-3)}{22x-69}},
\end{equation}
which depends only on \(\theta_{\rm i}\) and \(x\), and where $s_0 \equiv s(T_0) \simeq 2.9 \times 10^3$~cm$^{-3}$, $\rho_\text{crit} \simeq 1.1 \times 10^{-5}\, h^2$~GeV/cm$^3$ is the critical energy density of the Universe, $h \simeq 0.674$, and $\Omega_\text{DM} \simeq 0.26$~\cite{Planck:2018vyg}.
The value of \(\Tend\) that corresponds to this minimum can be obtained in a similar way and is fixed by the above relation. Notice that this relation is independent of \(\Tc\) as long as it is below \(\Tqcd\), therefore, for a given value of \(x\) the same minimum mass can be reached by a range of histories spanning \(\Tbbn < \Tc < \Tqcd\). This also means that the same minimum mass can be achieved in histories with different amounts of dilution from entropy injection. The range of the \(x\)-dependent exponent is between \(2/5\) to \(6/11\) for
\(x < -3\).
Thus, plugging in values for the known quantities in Eq.~\eqref{eq:ma_minimum} we can find the smallest possible axion mass range that reproduces the DM abundance in the three-crossing regime 
\begin{equation}
    m_a\gtrsim
    \begin{dcases}
        5\times 10^{-9}\, \mbox{eV}\, \left(\frac{\theta_{\rm i}}{0.5}\right)^{4/5} &\text{ for } x = -3\,, \\
        5\times 10^{-8}\, \mbox{eV}\,\left(\frac{\theta_{\rm i}}{0.5}\right)^{12/11} &\text{ for } x \rightarrow -\infty\,,
    \end{dcases}
\end{equation}
where we have used Eq.~\eqref{eq:ma_fa} for the product \(f_a\, m_a\). 
We should note that when \(x\) is equal or very close to \(-3\), the cosmology no longer supports three distinct crossings, but rather an extended period of $3\,H=m$. However, for $x\lesssim-3$, where three crossing are supported, the minimum mass remains close to the value shown above ($m_a\gtrsim 7\times 10^{-9}$~eV for $x=-4$). 

Interestingly, roughly the same minimum mass can also be obtained for a history that only experiences one crossing. If \(\Tend < \Tqcd\), then the entire nonadiabatic phase of increasing temperature occurs at temperatures below \(\Tqcd\), rendering the axion mass constant throughout this period. In this case, it is not possible to have multiple crossings because the mass does not experience the altered temperature evolution. However, the correct DM abundance can still be established for axion masses that are roughly as small as the minimum for three-crossing scenarios (cf. left side of Fig.~\ref{fig:m_Tc}) even though only one crossing occurs. 
The minimum possible mass for this kind of one-crossing scenario can be found analogously to Eq.~\eqref{eq:ma_minimum} and is given by:
\begin{equation}
    m_a \gtrsim \sqrt{\Tqcd\, \Tc} \left(\frac{M_{\rm P}}{ \Tqcd}\right)^{\frac{3+2x}{3\, (2x-5)}} \left(\frac{45\, (f_a\, m_a)^2\, \theta_{\rm i}^2\, s_0}{40\, \Omega_{\rm DM}\, \rho_{\rm crit}\, M_{\rm P}^2\, \Tqcd}\right)^{\frac{2\, (2x-3)}{3\, (2x-5)}}.
\end{equation}
Note that, in contrast to the three-crossing case, this expression does have a dependence on \(\Tc\), in addition to \(\theta_{\rm i}\) and \(x\), therefore the smallest value is reached only when \(\Tc \approx \Tbbn \approx 4\)~MeV. As before, the value of \(\Tend\) which corresponds to this minimum mass is fixed by the expression above. The range of the two \(x\)-dependent exponents (from left to right) is between \(1/11\) to \(1/3\), and \(6/11\) to \(2/3\), for \(x < -3\), respectively. 
Plugging in numbers we have 
\begin{equation}
    m_a\gtrsim
    \begin{dcases}
        8\times 10^{-9}\, \mbox{eV}\,\left(\frac{\Tc}{4\,\mbox{MeV}}\right)^{1/2} \left(\frac{\theta_{\rm i}}{0.5}\right)^{12/11} &\text{ for } x = -3\,, \\
        5\times 10^{-8}\, \mbox{eV}\,\left(\frac{\Tc}{4\,\mbox{MeV}}\right)^{1/2}\left(\frac{\theta_{\rm i}}{0.5}\right)^{4/3} &\text{ for } x \rightarrow -\infty\,.
    \end{dcases}
\end{equation}
This range is essentially the same as the range for three crossings, especially considering corrections when \(\gs(T)\) is included. 

\section{Axion Coupling to two Photons} \label{sec:photons}
With the full range of axion masses for the correct DM abundance understood, we now turn to detection prospects in our scenarios. The axion coupling to two photons is one of the most exploited interactions used to look for signatures in observations and experimental searches. The interaction takes the form
\begin{equation}
    \mathcal{L} = -\frac14\, g_{a \gamma}\, a\, F_{\mu \nu} \tilde F^{\mu \nu} = g_{a \gamma}\, a\, \vec E \cdot \vec B\,,
\end{equation}
where the coupling constant $g_{a \gamma}$ is model dependent and is related to the PQ scale $f_a$ as
\begin{equation}
    g_{a \gamma} = \frac{\alpha}{2\pi\, f_a} \left(\frac{E}{N} - \frac23\, \frac{4 + z}{1 + z}\right) \simeq 10^{-13}~\text{GeV}^{-1} \left(\frac{10^{10}~\text{GeV}}{f_a}\right),
\end{equation}
where $z \equiv m_u/m_d$ and $E$ and $N$ are the electromagnetic and color anomalies associated with the axion anomaly. For KSVZ models $E/N = 0$~\cite{Kim:1979if, Shifman:1979if}, whereas for DFSZ models $E/N = 8/3$~\cite{Dine:1981rt, Zhitnitsky:1980tq}. 

\begin{figure}
	\centering 
	\includegraphics[scale=0.40]{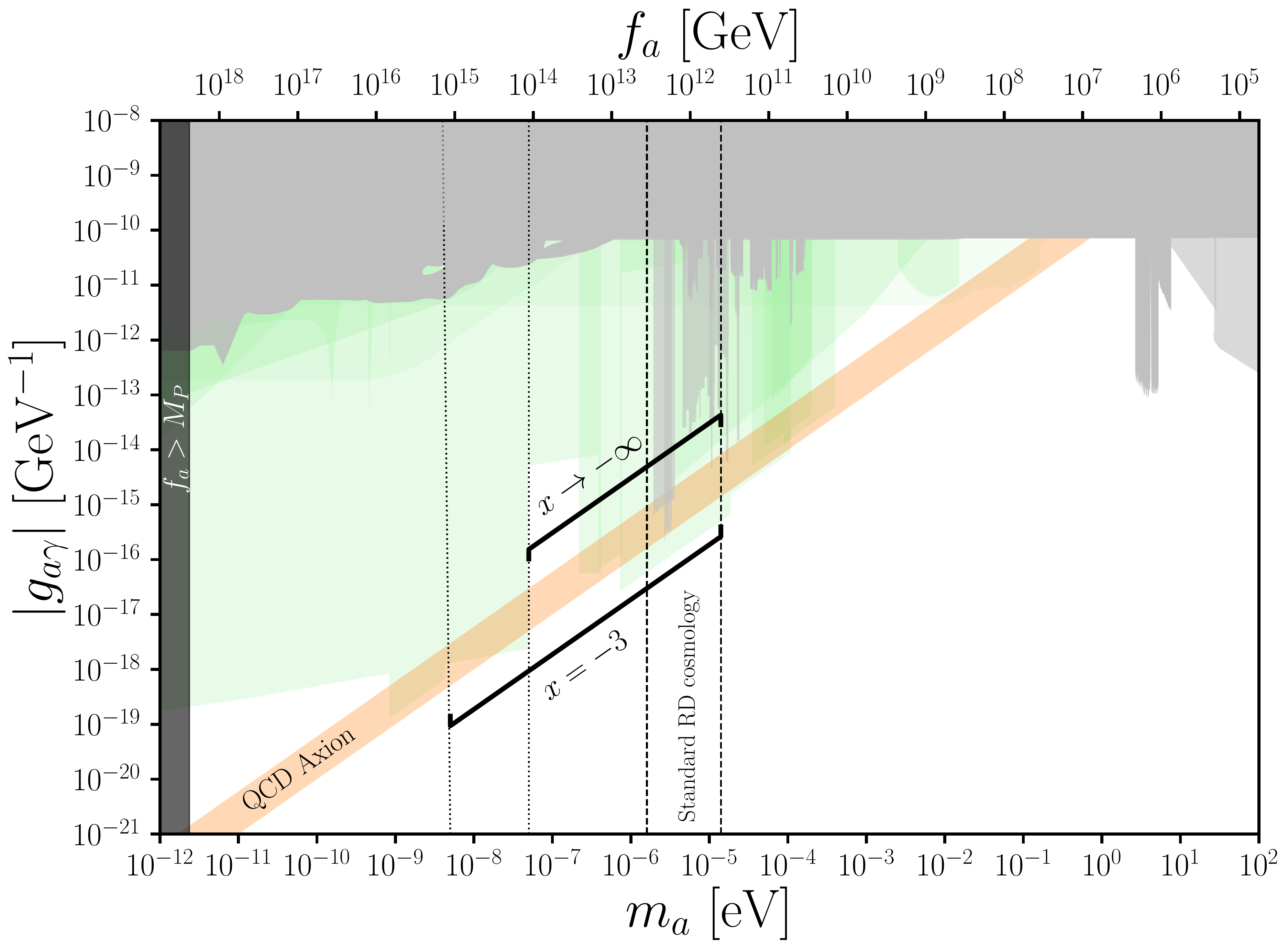}
    \caption{The axion parameter space for the axion-photon coupling in the standard RD cosmology as well as in the NSC scenarios considered here. The extent in mass of our NSC scenarios is shown for two benchmark cases: $x = -3$ and $x \to -\infty$, both in the three-crossing regime. The gray shaded areas correspond to excluded parameter space from cosmology, astrophysics, and laboratory experiments. Green shaded areas show prospects in sensitivity from various experiments. Figure adapted from Ref.~\cite{AxionLimits}.}
    \label{fig:ga_NSC}
\end{figure}
In Fig.~\ref{fig:ga_NSC} we show the parameter space of QCD axions (orange band) and the exclusion bounds from several astrophysical observations, cosmology-based constraints, and laboratory searches (gray shaded area)~\cite{Billard:2021uyg}. We also show several experimental concepts (green shaded area), with corresponding projections of sensitivity which emphasize coverage of the QCD axion band in particular, see e.g. Ref.~\cite{Adams:2022pbo}. The standard axion DM window extends between the vertical dashed lines, 
where the relic abundance is obtained from the misalignment mechanism in a standard RD cosmology with initial angles in the range $\theta_\text{i} \in [1/2,\, \pi/\sqrt{3}]$, such that both pre- and post-inflationary scenarios are included without invoking a fine-tuned solution. For the same range of initial angles, we show the mass range in which the DM relic abundance can be obtained in our scenarios for the two limiting values of the parameter \(x\) that support three crossings: \(x = -3\) and \(x \to -\infty\). Our extended mass window overlaps with the standard window on the high-mass end and reaches down to $m_a \approx 5\times 10^{-9}$~eV on the low end, where the extent of the lowest mass is determined by \(x\). 
This low-mass extension of the axion DM window overlaps with future projections which target the parameter space both within and outside of the standard window, allowing our NSC scenarios to be tested as this space becomes increasingly explored. 
For instance, several currently operating and proposed haloscopes aim to explore masses between the $\mu$eV to meV range, such as CULTASK~\cite{CAPP:2020utb, Lee:2020cfj} KLASH~\cite{Alesini:2017ifp}, ORGAN~\cite{McAllister:2017lkb, McAllister:2020twv}, RADES~\cite{Melcon:2018dba, CAST:2020rlf} and QUAX~\cite{Alesini:2020vny}. A related concept makes use of dielectric plates, such as the MADMAX~\cite{MADMAX:2019pub} experiment. The LC circuit based detector~\cite{Sikivie:2013laa} is expected to reach axion masses in the $10^{-7}$~eV to $10^{-9}$~eV range, while the ABRACADABRA experiment~\cite{Kahn:2016aff, Ouellet:2018beu} and ADMX SLIC~\cite{Crisosto:2019fcj} have already released promising results. Lastly, topological insulators have the potential to explore masses in the meV range~\cite{Schutte-Engel:2021bqm}.

\section{Conclusions}
\label{sec:concl}
In this work, we have considered the axion misalignment mechanism in a nonstandard history of the Universe that involves a period of matter domination by a field that decays with a time-dependent decay rate, parameterized as $\Gamma(R) \propto (\Rend/R)^x\, H_{\rm fin}$. The thermal history of the Universe then features a period of increasing temperature toward the end of the nonstandard cosmology (NSC) for \(x < -3/2\), before recovering the usual decrease once radiation domination is restored. This period of increasing temperature, in turn, significantly affects the evolution of the axion field, due to the temperature dependence of the mass of the axion at temperatures higher than \(\Tqcd\), and results in the possibility of restoring Hubble friction in the axion dynamics if \(x < -3\), {\it i.e.} introducing additional crossings (for up to three in total) of the axion mass and the Hubble expansion rate during its cosmological history. 

We have defined the conditions needed for a history to support three crossings (happening when the scale factor $R=R_1$, $R=R_2$, and $R=R_3$) and have explored the resulting parameter space both analytically and numerically. We have also performed a detailed numerical study of the axion dynamics in such nonstandard histories. 
We find that the axion evolution, and thus the misalignment mechanism itself, are significantly affected by a period of increasing temperature, most notably when three crossings occur. When a total of only one crossing occurs in the axion's history, the situation is largely unchanged from the standard radiation-dominated (RD) case or from a typical early matter dominated period with constant decay rate, depending on when the single crossing occurs. However, when multiple crossings occur (two or three), the axion field can experience substantial delays in the onset of oscillation (as in the case of two crossings where \(R_1 = R_2\)) or even periods during which Hubble friction is restored and oscillations temporarily stop (as in the case of three crossings, as well as two crossings where \(R_2 = R_3\)). 
Finally, there is also the possibility of an extended period during which \(3\, H(R) = m(R)\), instead of individual discrete crossings, though this situation is rather tuned, requiring essentially a single value of the axion mass \(m_a\) in order to be realized. 

In histories that have experienced three crossings, Hubble friction is temporarily restored in the period between the second and third crossings as the axion mass has fallen below the Hubble rate due to its temperature dependence. This restoration results in a kind of resetting of the initial conditions of the misalignment mechanism such that a second phase of oscillation proceeds from a new configuration of the axion field after the third crossing, with a smaller value of the misalignment angle. Furthermore, due to the generation of kinetic energy during the first phase of oscillations, the evolution of the axion between \(R_2 < R < R_3\) proceeds with notable differences based on how its potential and kinetic energies behave, with potential energy density remaining constant up to changes in the axion mass, and kinetic energy density redshifting as \(R^{-6}\). 

We find that the resulting axion energy density can account for the entire observed DM abundance today for a wide range of axion masses, extending from the standard-history window of $10^{-6}$~eV $\lesssim m_a\lesssim 10^{-5}$~eV down to a minimum mass between $m_a \approx 5\times 10^{-9}$~eV to $5\times 10^{-8}$~eV for \(x < -3\). In such increasing-temperature NSC scenarios, the axion mass for the correct DM abundance is always less than or equal to the standard RD case, due to strong entropy injection from the decay of the dominating field, as well as the smaller misalignment angle from which the final phase of oscillation proceeds. 

As mentioned in the introduction, other scenarios can also extend the axion window to smaller masses. If the axion is coupled to some other new fields into which it can decay, a smaller mass can compensate for the depleted abundance~\cite{Agrawal:2017eqm, Allali:2022yvx}. Such scenarios can be distinguished from ours through potential signatures in experiments related to the additional fields or through the presence of dark radiation. On the other hand, trapped misalignment scenarios, in which the axion is temporarily trapped in the wrong minimum, also induce two periods of oscillation and extend the mass window to smaller values~\cite{DiLuzio:2021gos, DiLuzio:2021pxd}. The parameter space in which the DM abundance is reproduced in such models places the axion outside of the QCD axion band, thus, a potential discovery would clearly differ from our scenario in this case. However, when it comes to other NSC scenarios, the problem of multiple possible histories leading to the same observables is more pronounced. For example, our extended mass window for histories with periods of increasing temperature overlaps with the window corresponding to matter-like NSC histories with the typical constant decay rate~\cite{Arias:2021rer} in a way that cannot be distinguished with axion observables alone. One potential direction that may help break such degeneracies is the imprint of NSC histories on the spectrum of a potential gravitational wave background. From the shape of the spectrum, one can reconstruct the expansion history and therefore access information about the NSC period itself. We leave investigations along these lines for future work.

Finally, we have mapped the mass range accessible to our NSC scenarios onto the exclusion plot for the axion coupling to two photons. 
Importantly, our mass range extends to axion masses significantly smaller than the standard window, and overlaps with future projections that lie outside of the standard range. This allows for scenarios such as ours, and thus modifications to the cosmological history before BBN, to be probed in the coming years, and lends increased motivation to continued searches beyond the standard window. 

\section*{Acknowledgments}
NB thanks Universidad de Santiago de Chile for its hospitality during the completion of this work.
PA acknowledges funding from FONDECYT project 1221463.
NB received funding from the Patrimonio Autónomo - Fondo Nacional de Financiamiento para la Ciencia, la Tecnología y la Innovación Francisco José de Caldas (MinCiencias - Colombia) grants 80740-465-2020 and 80740-492-2021.
NB is also funded by the Spanish FEDER/MCIU-AEI under grant FPA2017-84543-P.
JO and LR are supported by the project AstroCeNT: Particle Astrophysics Science and Technology Centre, carried out within the International Research Agendas programme of the Foundation for Polish Science financed by the European Union under the European Regional Development Fund.
This project has received funding and support from the European Union's Horizon 2020 research and innovation programme under the Marie Sk{\l}odowska-Curie grant agreement No.~860881 (H2020-MSCA-ITN-2019 HIDDeN).
This publication is based upon work from COST Action COSMIC WISPers CA21106, supported by COST (European Cooperation in Science and Technology).

\bibliographystyle{JHEP}
\bibliography{biblio}

\end{document}